\documentclass[5p]{elsarticle}
\usepackage{graphicx}
\usepackage{amssymb}
\usepackage{esvect}
\usepackage{aas_macros}

\title{Detectors for high-energy messengers from the Universe}
\author{W. Hofmann}
\author{J. Hinton\corref{cor1}}
\cortext[cor1]{Corresponding author: jim.hinton@mpi-hd.mpg.de}

\address{Max-Planck-Institut f\"ur Kernphysik, Postfach 103980, 69029 Heidelberg} 

\begin{document}
\begin{abstract}
  High-energy messengers from the Universe comprise charged cosmic rays, gamma rays and neutrinos. Here we summarize the detection principles and detection schemes for these particles, with a focus on ground-based instruments which employ natural media such as air, ice, or water as their detection medium.
\end{abstract}

\maketitle


High energy astrophysics is concerned with the study of non-thermal
particle populations in our Galaxy and beyond, with their sources,
propagation, and impact on their cosmic environment.
This field
relies on detectors for high-energy messengers from the
Universe -- the subject of this article -- but also on astrophysical
instruments in many other domains of the electromagnetic spectrum, most notably in the radio and
X-ray, tracing the synchrotron radiation of high energy electrons. For
the current discussion, we will -- somewhat arbitrarily -- concentrate
on the domain from GeV energies up, where the detectors address common science
themes and share many detection features; MeV instruments differ in
terms of their science focus but also in their detection principles.

\section{High Energy Messengers}

\subsection{Cosmic Rays} 

\begin{figure}[htbp]
\begin{center}
\includegraphics[width=0.47\textwidth]{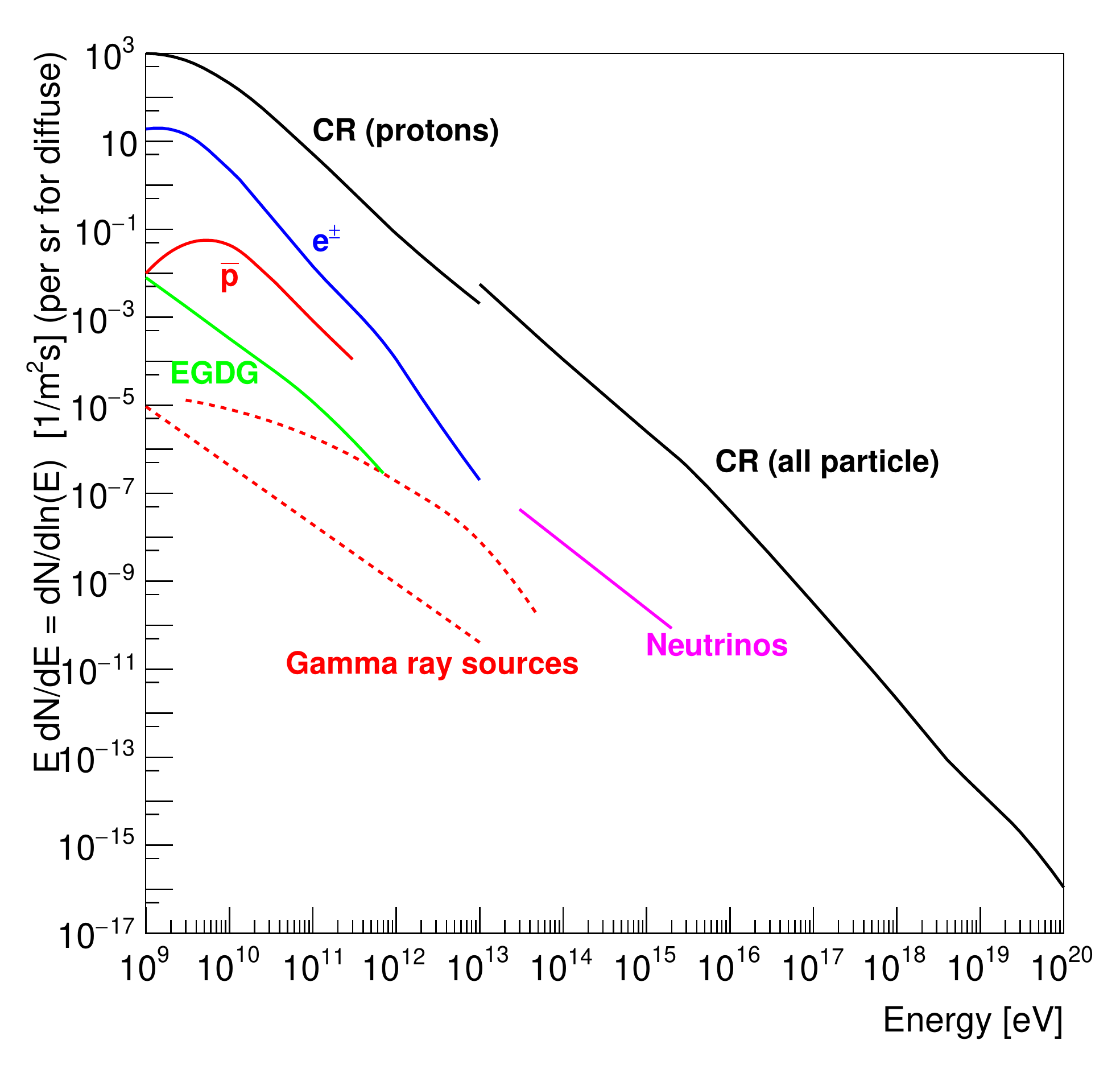}
\caption{The flux of various cosmic messengers expressed as the rate per logarithmic energy interval $E dN/dE = dN/d\ln(E)$. 
For diffuse and close to isotropic fluxes, a solid angle of 1~sr is used. 
Indicated are the diffuse all-particle cosmic-ray flux \cite{2009PrPNP..63..293B} resp. the proton flux (black)
 \cite{2015PhRvL.114q1103A}, the antiproton flux (red) \cite{Adriani:2017bfx,Aguilar:2016kjl}, the flux of cosmic e$^\pm$ (blue) 
 \cite{Aguilar:2014mma,Ambrosi:2017wek}, the flux
of extragalactic diffuse gamma rays (EGDG, green) \cite{2015ApJ...799...86A}, and the high-energy neutrino flux (magenta) \cite{2017arXiv170103731I}. 
Also shown are the gamma-ray fluxes from a localized
strong source (RX\,J1713.7-3946, \cite{2016arXiv160908671H}) and from a faint source (NGC\,253, \cite{2012ApJ...757..158A}) (red dashed). Assuming that these gamma rays are produced by hadronic processes, the flux of neutrinos from these sources is approximately equal to the gamma ray flux. Event numbers per spectral interval $\Delta \ln(E) = 1$ are obtained by multiplying the flux with the exposure (effective detection area $\times$ time $\times$ solid angle of the detector (for diffuse fluxes)). For cosmic ray detection, for example the PAMELA spectrometer provides an exposure of about 20 cm$^2$sr $\times$ 10~yr  $\approx 6 \times 10^5$ m$^2$ sr s; at the other end of the range, the Pierre Auger air shower detector has accumulated $\sim3 \times 10^{18}$  m$^2$ sr~s. For gamma ray detection of a given source, the time-integrated exposure of Fermi-LAT is of order $2 \times 10^7$ m$^2$s, and current Cherenkov telescope arrays provide up to $\approx 10^5$ m$^2$ $\times$ a few 100~h $\approx$ $10^{11}$ m$^2$s.
}
\label{fig_cr}
\end{center}
\end{figure}

Cosmic rays (e.g. \cite{2009PrPNP..63..293B, 1983ARNPS..33..323S}) -- highly relativistic charged particles of astrophysical origin --  impacting on Earth are
by far the longest established high energy messengers from the Universe. The measured quantities characterizing cosmic rays are the energy spectrum, their (mild) anisotropy, their elemental composition, and a potential temporal variability. Cosmic rays cover a huge range in energy -- up to $10^{20}$ eV -- and in flux, see Fig. \ref{fig_cr}. Apart from the low-energy region influenced by the modulation by the solar wind, the cosmic ray spectrum follows a power law in energy, $dN/dE \sim E^{-\alpha}$, with an index of $\alpha \approx 2.7$ up to the cosmic ray knee at about $10^{16}$ eV, where the spectrum steepens slightly, to flatten again at the ankle around $10^{18}$ eV, and cuts off around $10^{20}$ eV. The elemental composition coarsely resembles the solar system composition, dominated by protons, with notable enhancements for certain elements and isotopes that are rare in the solar system \cite{Simpson:1983hz}; in the region of the knee, the composition shifts towards heavier nuclei \cite{Kampert:2012mx}.  Cosmic rays arrive nearly isotropically, with energy-dependent anisotropies on various angular scales at the level up to $10^{-3}$ \cite{Deligny:2016asd}. Temporal variability of cosmic rays is primarily induced by the solar wind following the solar cycle.

The standard paradigm for origin of Galactic cosmic rays (see e.g. \cite{2013A&ARv..21...70B}) -- theoretically plausible but not experimentally verified in full -- is that cosmic rays at least up to the knee, possibly up to the ankle are accelerated in our Galaxy via Fermi acceleration mainly in supernova remnants.
Cosmic ray composition reflects the composition of ambient material injected into the acceleration process, modified by the volatility of materials which might be locked up in grains of interstellar dust. Cosmic rays then propagate diffusively, with a (poorly known) diffusion coefficient $D$ of a few $\times 10^{28} E_{\mathrm{GeV}}^{0.3...0.6}$ cm$^2$/s,
\footnote{The notation for units is that $E_{\mathrm{GeV}}$ stands for energy in units of GeV, etc.}
 see e.g. \cite{2007ARNPS..57..285S}.
With a gyroradius $r_{g,\mathrm{pc}} \sim E_{\mathrm{PeV}}/B_{\mathrm{\mu G}}$
and scattering length $\lambda \sim D/c \sim E_{\mathrm{GeV}}^{0.3...0.6}$ pc, short compared to galactic distance scales even at PeV energies and beyond, cosmic rays are almost perfectly isotropized, with small anisotropies reflecting (a) the net diffusive flow governed by a non-uniform distribution of sources, and (b) the local structure and degree of turbulence of interstellar magnetic fields. Energy-dependent escape from the disk of the Galaxy steepens the spectrum by $\Delta \Gamma \approx 0.5$. During propagation, elemental and isotopic composition are modified by spallation processes; from the composition the amount of matter traversed and the residence time of cosmic rays in the Galaxy can be deduced; at GeV energies, cosmic rays travel $O(10^7)$ years before escaping from the Galaxy. The highest-energy cosmic rays are believed to be of extragalactic origin; the number of objects able to confine cosmic rays during acceleration up to $10^{20}$ eV -- requiring $r_{g}$ to be small compared to the size of the acceleration region -- is very limited and includes e.g. the giant radio lobes of active galaxies (see e.g. \cite{1984ARA&A..22..425H, 2011ARA&A..49..119K}).

Antiparticles \cite{2017arXiv170906507B} among cosmic rays -- positrons, antiprotons and also light anti-nuclei -- have received significant attention since they probe another potential source of cosmic ray particles: the (charge-symmetric) annihilation or decay of heavy relics from the Big Bang, such as weakly interacting dark matter particles \cite{Cirelli:2010xx}. Fig. \ref{fig_cr} indicates the spectrum of antiprotons as well as of electrons plus positrons; the latter falls like $E^{-3}$ up to about 1~TeV, where it exhibits a rather sharp knee; the fraction of positrons rises with energy up to a peak around 1 TeV.  Alternative explanations of antiparticle origin include interactions during cosmic-ray propagation, 
and production in cosmic-ray sources (with positrons being pair-produced in the vicinity of pulsars, 
or injected as products of $\beta$ decays of freshly produced heavy elements in supernova explosions). 
Spectra and anisotropy may ultimately allow the, currently heavily debated, origin of antiparticles to be pinned down.
Direct cosmic ray measurements probe only local cosmic rays, and -- except possibly for the $10^{20}$ eV domain where deflection is modest \cite{2011ARA&A..49..119K} -- do not allow imaging of the sources in an astronomical sense. Observations of the undeflected cosmic gamma rays and neutrinos therefore provides complementary information.

\subsection{High Energy Gamma Rays and Neutrinos} 

High energy gamma rays and neutrinos are created when cosmic rays interact with ambient matter and radiation fields. These neutral messengers propagate on straight paths and thereby allow imaging of the (projected) cosmic ray density in the Universe, the identification of cosmic ray sources as regions of peak cosmic ray density, and the study of cosmic ray propagation as these move away from their sources. Gamma rays and neutrinos arriving at Earth are characterized by their angular distribution -- the sky image -- and by their energy spectrum. While the Universe is practically transparent for neutrinos, this is not the case for gamma rays: the primary mechanism of absorption is the interaction with low-energy target ($E_T$) photons such as starlight; the threshold for pair production is $E_{\gamma, \mathrm{TeV}} \approx 1/E_{T, \mathrm{eV}}$. Compact sources with strong local radiation fields -- such as pulsar magnetospheres or AGN-- may be opaque to the gamma rays produced in them. During propagation in intergalactic space, absorption on background light limits the gamma ray mean free path at 100 TeV to a redshift $z \approx 0.001$, at 1 TeV to $z \approx 0.1$, and at 100~GeV to $ z \approx 1$ \cite{2017A&A...603A..34F}. Another difference is that for gamma rays, a second production mechanism is important: the up-scattering of ambient photons by high-energy electrons and positrons, frequently resulting in an ambiguity regarding the nature of the parent population. Photo-production of neutrinos is possible, but should not play a significant role for most environments and energy domains.

For gamma rays produced in proton interactions (for simplicity neglecting heavier nuclei among cosmic rays), in the case of a power-law distribution of protons, $dN/dE \sim E^{-\Gamma_p}$, the secondary gamma rays also follow a power law with almost the same index, $\Gamma_\gamma \approx \Gamma_p$ (with small deviations due to the slight energy dependence of proton interaction cross sections and scaling deviations in inclusive particle production) \cite{2006PhRvD..74c4018K}; as a more general rule of thumb, the gamma ray spectrum at energy $E_\gamma$ reflects the proton spectrum at $E \approx 10 E_\gamma$, although structure or cutoffs in the proton spectrum appear smoothed in the gamma-ray spectrum \cite{2006PhRvD..74c4018K}. The gamma ray luminosity of a region is given by $L(E_\gamma>E_o)\approx Q(10 E_o)/\tau$ where $Q(E)$ is the energy in protons above energy $E$ contained in the region. The effective interaction time $\tau_\mathrm{sec} \approx 5 \cdot 10^{15}/n_{\mathrm{cm}^{-3}}$ reflects  the local density $n$ of target gas. Corresponding relations apply for neutrino production, except that the energy fraction going into neutrinos (mostly via charged-pion decays) is slightly smaller than the energy fraction going into gamma rays (mostly via neutral pion decays) \cite{2006PhRvD..74c4018K}, and that for neutrino-flavor-specific detection the flavor composition and neutrino mixing during propagation needs to be considered.

For gamma-ray production by electron primaries, the typically dominant radiation process is inverse Compton up-scattering of ambient photons. In a radiation field of energy density $U$, the energy loss time scale is given by $\tau_{\mathrm{yr}} = E_e/(dE_e/dt) \approx 3 \cdot 10^{5} U^{-1}_{\mathrm{rad,eV cm^{-3}}} E^{-1}_{e,\mathrm{TeV}}$ (as long as scattering proceeds in the Thompson regime) 
\cite{1970RvMP...42..237B} \footnote{Given that typical Galactic radiation fields provide $U \approx 1$ eV/cm$^3$, the lifetime of cosmic-ray electrons is limited to
$\tau_{yr} \approx 3 \cdot 10^{5}/E_{e,\mathrm{TeV}}$ and hence their range to $ \sqrt{D \tau} \approx 1 \mathrm{kpc}/E_{e,\mathrm{TeV}}^{0.25}$.} . Since the scattering cross section depends on energy, the spectral index of gamma rays is given by $\Gamma_\gamma = (\Gamma_e+1)/2$.

Fig. \ref{fig_cr} illustrates the gamma ray flux measured from a strong source and from a faint source near current detection limits. The interpretation of these gamma ray fluxes is complicated by the facts that (a) the target density is frequently poorly known, directly translating in a corresponding uncertainty in the measurement of $Q(E)$, and (b) it is frequently difficult to distinguish between a hadronic and electronic origin of the gamma rays, which -- due to the vastly different interaction time scales of nuclei and electrons -- translates into order-of-magnitude uncertainties in $Q(E)$. For example, the gamma ray spectra of the remnant RX\,J1713.7$-$3946 shown in Fig. \ref{fig_cr} can be modeled as originating from a $6 \cdot 10^{49}$ erg proton population, or a $10^{47}$ erg electron population \cite{2016arXiv160908671H}. Fig. \ref{fig_cr} also illustrates the high-energy neutrino flux measured by IceCube \cite{2017arXiv170103731I}; the neutrino flux appears to be diffuse, and no consensus exists regarding the exact (but likely extragalactic) origin of these neutrinos.

\section{Detection Characteristics and Requirements}
\label{sec_det}

\subsection{Effective Detection Area, Rates, Resolution, Particle Identification}

A detector for high energy messengers is usually characterized by an effective detection area, angular resolution in reconstructing the direction of the incident particle, energy resolution and -- in the case of detectors for charged particles -- its ability to identify different particles and to determine their mass and/or charge.

The effective detection area is given by
$$
A_{\rm eff}(E) = \int \epsilon(\vv{r}, E) d^2r 
$$
where $\epsilon(\vv{r})$ is the probability to detect a particle of energy $E$ whose trajectory impacts at $\vv{r}$ in a suitably chosen detector plane. The detection rate $R(E)$ is given by the product of particle flux $\phi(E) = dN/dtdE$ and effective area, $R(E) = \phi(E) A_{\rm eff}(E)$. For instruments with wide apertures detecting particles over a wide range of angles of incidence (as opposed to a parallel stream of particles from a source at large distance), often the effective area $\times$ solid angle product is used
$$
A_{\rm eff}(E) = \int \epsilon(\vv{r}, \Omega, E) d^2r d\Omega
$$
and the detection rate is $R(E) = dN/dtdEd\Omega A_{\rm eff}(E)$.

Charged cosmic rays and gamma rays have relatively short interaction lengths of the scale of 10s of g/cm$^2$; charged cosmic rays leave ionization tracks in matter. Hence for most detector types the particles have a unit probability of interacting  provided that they intersect the detector (as opposed to passing through the detector without leaving any energy deposition), and the efficiency $\epsilon$ essentially refers to the probability of depositing sufficient energy and information in the sensitive elements of the detector, to register the particle and reconstruct (some of) its properties. The situation differs for neutrino detectors, where the typical mean free path of a neutrino (in water) is $2 \times 10^9$ m at $10^{12}$ eV,  $2 \times 10^7$ m at $10^{15}$ eV and
 $10^6$ m at $10^{18}$ eV; here $\epsilon$ can be coarsely factorized in to an interaction probability $\epsilon_{int}$ depending on the total depth of the detector, and the probability $\epsilon_{reg}$ of registering the interaction.

For cosmic ray and gamma ray detectors the probability of interacting ahead of the detection volume is usually very small; for neutrino detection absorption may be significant, for example for upward-going very high energy neutrinos that have to traverse the Earth.

In general, instruments will have a finite energy resolution and angular resolution, and possibly misidentify particles. The finite energy resolution (described by a suitable migration matrix $K$) results in a detection rate 
$$
R(E_{\rm det}) = \int \phi(E) A_{\rm eff}(E)  K(E_{\rm det},E) dE
$$ 
and is usually addressed either by fitting spectral models and taking energy migration into account, or by applying suitable deconvolution methods, which usually make assumptions regarding the smoothness of the intrinsic spectrum, to regularize the deconvolution and avoid cumulation of statistical errors.

The requirements for exposure (defined as effective detection area $\times$ integration time) can be read from Fig. \ref{fig_cr}; for cosmic rays in the GeV domain exposures of order 1 m$^2$ s sr are ample, whereas around $10^{20}$ eV exposures well beyond $10^{17}$ m$^2$ s sr are required. 
Energy resolution is in general uncritical given that typically spectra are power laws with smoothly varying spectral index. An energy resolution 
$\Delta E/E \approx 0.1...0.2$ is usually enough to follow features or cut-offs in spectra with sufficient fidelity. There is one primary application that benefits from enhanced energy resolution: the search for gamma ray lines or line-like features created in the annihilation of dark matter particles. For charged cosmic rays (that are anyway deflected in interstellar magnetic fields), angular resolution is rather uncritical; only the search for peaks in the arrival direction of ultra-high-energy cosmic rays benefits from resolution in the degree range and below. Detectors for gamma rays and neutrinos benefit from high angular resolution, both in resolving structures within sources, and in terms of sensitivity for point sources. Whenever significant isotropic backgrounds exist, for example due to imperfectly rejected cosmic ray events, the detection sensitivity is given by $S \approx N_{\mathrm{signal}}/\sqrt{N_{\mathrm{BG}}}$, where $N_{\mathrm{BG}} \sim \epsilon_{\mathrm{BG}} \theta^2$ for a given angular resolution $\theta$ and background efficiency after cuts  $\epsilon_{BG}$, and hence  $S \sim \epsilon_{BG}^{-0.5}\theta^{-1}$.

\subsection{Detection Principles}

For apertures up to m$^2$, it is possible to fly detectors in space, on dedicated carriers or the International Space Station, or to use high-altitude balloons, with a typical remaining grammage on top of the detector of few g/cm$^2$. Such detectors directly register the primary particles. Particle detectors in space use identical techniques to laboratory particle detectors, in the case of charged-particle detectors often combining a magnetic spectrometer equipped with silicon strip tracking layers, time-of-flight counters, ring-imaging Cherenkov detectors, $dE/dx$ measurements, and transition radiation detectors for particle identification, and imaging calorimeters for energy measurement and electron-hadron separation. The main challenge in constructing and operating such systems lies in the mechanical loads during launch, the lack of access during operation, and the often very large temperature swings depending on the orientation relative to the sun.

At higher energies, space-based detectors cannot provide sufficiently large effective areas, hence above some (particle type-dependent) energy instruments must rely on using natural large-volume media as calorimeters, most notably the atmosphere (for cosmic rays and gamma rays) and water or ice (for neutrinos). A variety of mechanisms can be used to `read out' the calorimeter, including the ionization energy loss in (often sparse) particle detectors at ground level, the fluorescence and Cherenkov light emitted by shower particles, the radio emission of shower particles, or even the acoustic signal created by the localized heating of the medium due to shower energy deposition. 

The following discussion will focus primarily on this second type of detection system that is a characteristic tool of astroparticle physics; the technology of space-based spectrometers and calorimeters is very similar to that used in laboratory experiments and addressed elsewhere in this volume, apart from the specific challenges imposed by the space environment. For very detailed review of space-based detectors, the reader is referred to \cite{2014arXiv1407.7631B}; key instruments include e.g. PAMELA \cite{Adriani:2017bfx} and AMS02 \cite{Ting:2013sea} for cosmic rays and Fermi-LAT \cite{Atwood:2009ez} for gamma rays.

\section{Shower Characteristics and Detection Basics}
\label{sec_showers}

At higher energies, where space-based direct detection suffers from the limited detection areas, detection methods all deal with the detection of particle cascades induced either in the Earth's atmosphere (air showers) or in dense natural media. The single exception is the case of muon-neutrino detection, where the high energy muon produced in the first interaction typically has a very long track, and radiative losses initiating cascades are suppressed because of the large muon mass, starting to become relevant in the TeV domain \cite{Patrignani:2016xqp}.

\subsection{General properties of showers}

\begin{figure}[h]
\begin{center}
\includegraphics[width=9cm]{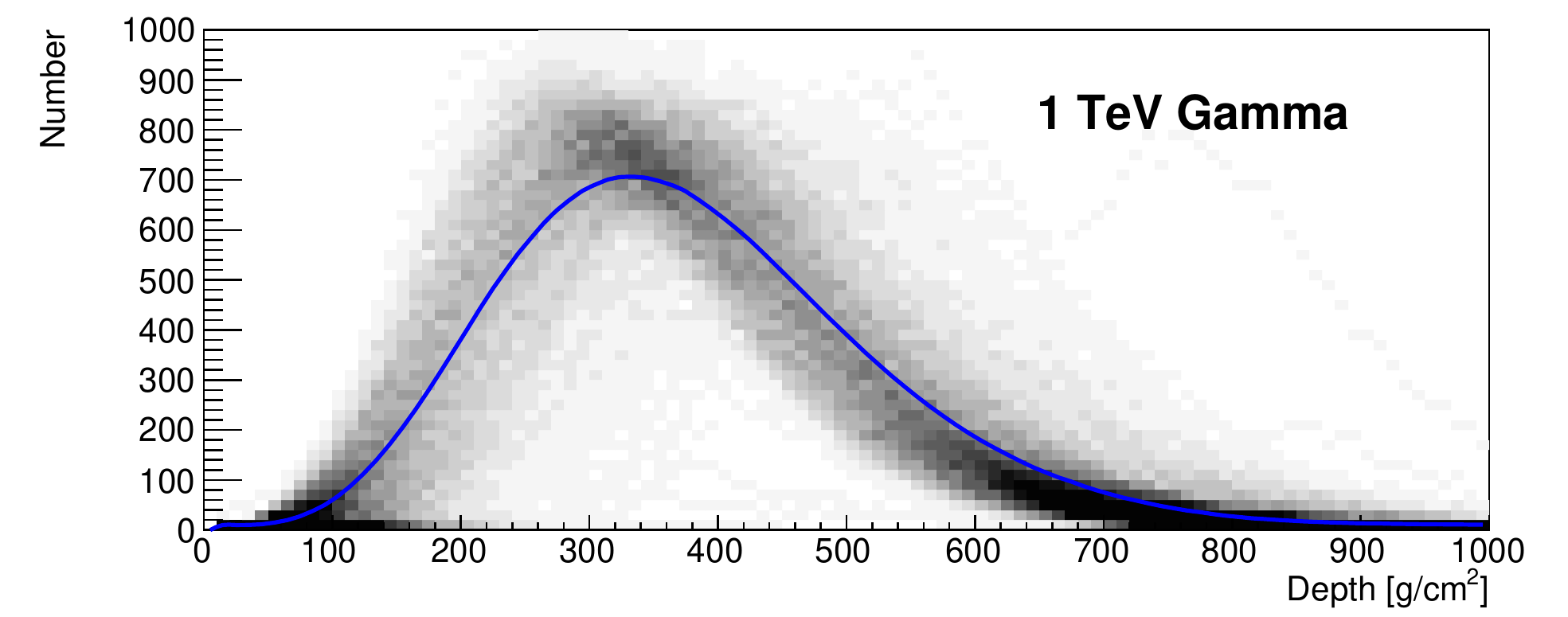}
\includegraphics[width=9cm]{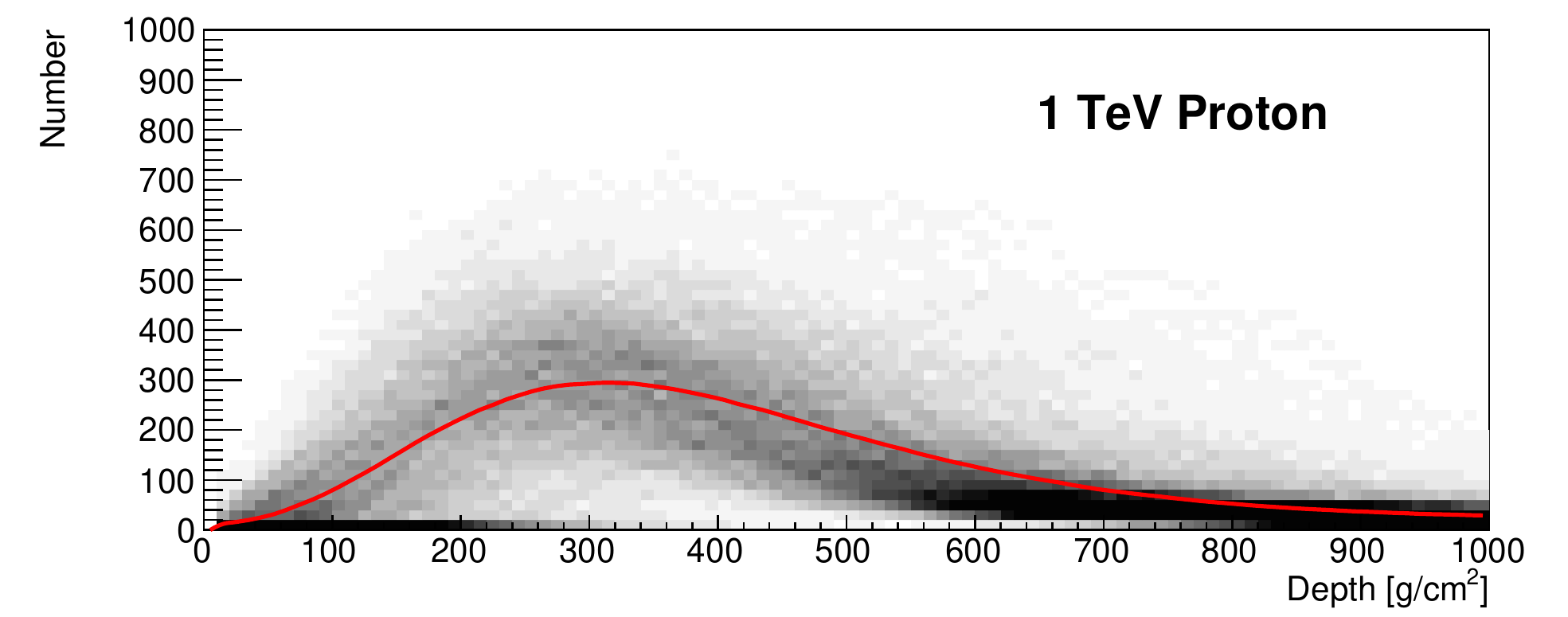}
\includegraphics[width=9cm]{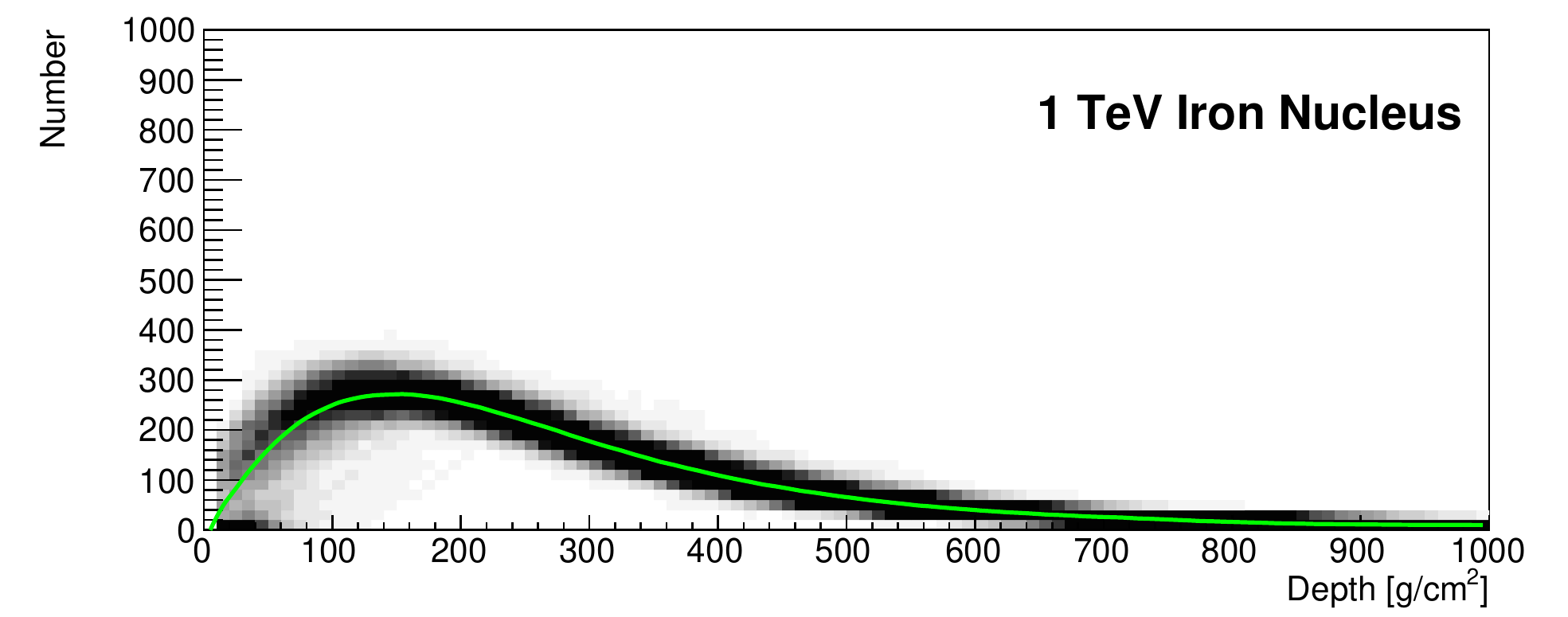}
\caption{Characteristics of air showers: number of charged particles as a function of depth; shown is the distribution (for 1000 showers) in the number of charged particles as a function of atmospheric depth, as well as their mean number, at 1 TeV primary energy, for gamma primaries (top), proton primaries (middle) and iron primaries (bottom). Based on CORSIKA simulations.}
\label{fig_sho1}
\end{center}
\end{figure}

In case of gamma-ray induced showers, the cross-sections of electromagnetic processes governing shower evolution are exactly known and  the basic properties of showers can be derived analytically \cite{Rossi:1941zza,1970RvMP...42..237B}. A simplified description is provided by the Heitler model where electrons, positrons and gammas undergo repeated symmetrical branchings, with electrons or positrons radiating a gamma, and gammas converting into electron-positron pairs. The characteristic splitting length $d$ over which an electron loses half its energy is related to the radiation length $\lambda_r$ of the medium via $d = \lambda_r \ln{2}$. After $n$ branchings, the total number of particles in the shower is $2^n$, and the energy per particle $E_0/2^n$. Multiplications stops when particles reach the critical energy $E_C$ below which ionization energy loss dominates. The shower maximum with $E_0/E_C$ particles is reached after $n = (1/\ln{2})\ln{(E_0/E_C)}$ splittings, at depth $X_{\rm max} = \lambda_r \ln{(E_0/E_C)}$. The elongation rate 
$\Lambda \equiv \mbox{d}X_{\rm max}/\mbox{d}\log{E_0}$ is in this model given by $2.3 \lambda_r$. Compared with shower simulations, this simple model
over-predicts the number of particles, as seen from Fig. \ref{fig_sho1} (top) where a 1 TeV gamma ray shower in air ($E_C \approx 87$ MeV) is shown, and it ignores the tail beyond the shower maximum, due to fluctuations in the lengths between branchings and the uneven sharing of energy. Profiles of electromagnetic showers (with depth $t$ measured in units of $\lambda_r$) can be parametrized by $dE/dt = E_0 b (bt)^{a-1}e^{-bt}/\Gamma(a)$ with $b$ varying from 0.7 for low-Z materials to 0.45 for high Z, and $a$ determined from the depth of shower maximum $X_{max}/\lambda_r = t_{\rm max} = (a-1)/b = \ln{(E_0/E_C)} + C_j$, with $C_e = 0.5$ for electron-induced cascades and $C_\gamma = +0.5$ for gamma-ray induced cascades \cite{Patrignani:2016xqp}. The transverse dimensions of showers scale with the Moliere radius 
$R_M \approx \lambda_r (21\,\mbox{MeV}/E_C)$; about 90\% of the energy is deposited inside $R_M$  \cite{Patrignani:2016xqp}. Table \ref{tab_sho1} lists characteristic dimensions of electromagnetic showers in water and air.

\begin{table}[htbp]
\begin{center}
\caption{Characteristic dimensions of electromagnetic showers in air ($\lambda_r \approx 36.6$ g/cm$^2$, $E_C \approx 87$ MeV, $R_M \approx 8.8$ g/cm$^2$) and in water ($\lambda_r \approx 36.1$ g/cm$^2$, $E_C \approx 78$ MeV, $R_M \approx 9.8$ g/cm$^2$). Height values for air showers refer to vertical incidence.}
\vspace{0.2cm}
\begin{tabular}{l c c c}
\hline
Energy [eV] & $10^{12}$ & $10^{15}$ &$10^{18}$ \\
\hline
$X_{\rm max} = \ln{(E_0/E_C)}$ & $\approx 9 \lambda_r$ &  $\approx 16 \lambda_r$ &  $\approx 23 \lambda_r$\\
in water [m] & 3.4 & 5.9 & 8.4 \\
in air [g/cm$^2$] & 340 & 600 & 850 \\
in air [km asl] & 8.3 & 4.2 & 1.5 \\
\hline
$R_M$ in water [m] & 0.10 & 0.10 & 0.10 \\
$R_M$ in air @ $X_{\rm max}$ [m] & 175 & 110 & 85 \\
\hline
\end{tabular}
\label{tab_sho1}
\end{center}
\end{table}

The description of  of hadronic cascades relies on empirical hadronic interaction models. The Heitler model
can be adapted to hadronic showers by assuming that after each interaction length $\lambda_I$, $N$ hadrons are produced, at each step with a certain fraction of energy diverted  -- via $\pi^0$ production -- into electromagnetic showers \cite{Matthews:2005sd,Montanus:2014lya}. The compared to $\lambda_r$ larger interaction length $\lambda_I$ is partly compensated by the larger multiplicity per interaction $N$, speeding up shower evolution.

Instrument design and data analysis rely very much on detailed simulations of showers, with GEANT4 \cite{Allison:2016lfl} as de-facto standard code for showers in dense media and for complex absorber geometries, and CORSIKA \cite{Heck:1998vt} for air showers, with options to handle Cherenkov emission and radio emission. The longitudinal development of simulated 1 TeV showers in air is illustrated in Fig. \ref{fig_sho1}, for different primaries. Compared to gamma primaries, nucleus-induced showers have only $\approx 1/3$ the number of charged particles, for same primary energy. At 1 TeV, longitudinal profiles do not differ strongly between proton showers and gamma air showers, except for a slower fall-off at large depth for protons, but proton showers show much larger fluctuations in shower evolution. Iron primaries, on the other hand, due to their large interaction cross section, result in showers that develop faster, and, due to the superposition of many nucleonic cascade, show rather small fluctuations from shower to shower.

Detectors for showers induced by very high energy cosmic messengers in general need to provide large effective areas, to detect the low particle fluxes; the large detector volumes exclude sampling calorimeters where absorber layers alternate with active layers detecting particles via their ionization energy loss. Detectors instead usually rely on transparent natural media such as air, water or ice, and detect shower particles via their emission of Cherenkov light, of fluorescence light (in air), or via their coherent radio emission, with sensors either at the borders of the medium, or sparsely distributed throughout the medium. 

The Cherenkov threshold for electrons in water is $E = 0.8$ MeV, the light yield (for $v \approx c$, in the 300 - 600 nm range) is $\approx 320$ photons/cm or $\approx 160$ photons per MeV energy loss, emitted at the Cherenkov angle of $41^\circ$. In air, Cherenkov threshold is height-dependent; for electrons, the threshold is $\approx 35$ MeV at 10 km asl and $\approx 20$ MeV at sea level, with Cherenkov angles of $\approx 0.8^\circ$ and $\approx 1.5^\circ$, respectively; the yield is 0.15 to 0.5 photons/cm or $\approx 500$ photons per MeV deposited by $e^\pm$ well above the Cherenkov threshold. 

In air showers, fluorescence light is isotropically emitted when charged shower particles excite $N_2$ molecules; emission is between 290 and 430 nm, with multiple lines, the strongest at 337 and 357 nm; the fluorescence yield is $\approx 7$ photons per MeV ionization energy loss \cite{Arqueros:2008cx}.  
Molecular nitrogen and oxygen also emit fluorescence light in the near infrared \cite{Conti:2010ux}, but atmospheric background light is high in this range of the spectrum.

Radio emission by showers  in dense media is attributed to the Askaryan effect \cite{1962JPSJS..17C.257A}, resulting from charge separation in the direction of shower propagation, since ionization electrons tend to move forward, leaving positive ions behind. The current associated with initially growing, and later dying out negative charge near the shower front leads to electromagnetic radiation. In air showers \cite{Huege:2016veh,Schroder:2016hrv}, a second effect contributes: the time-variable current created by the deflection and separation of shower electrons and positrons in the geomagnetic field. This mechanism creates emission with fixed linear polarisation, whereas for the Askaryan effect the polarisation points towards the shower axis; their relative strengths depend on the geomagnetic field and the orientation of the shower; the Askaryan effect is up to 20\% of the geomagnetic radiation. The two components interfere, resulting in a radio intensity that is not symmetrical around the shower axis. Unlike for other detection mechanisms, radio signals are created by coherent superposition of signals from individual shower particles -- at least for wavelengths larger than or comparable to the thickness of the shower front, in the range up to 10s of MHz for air showers -- and the radiated energy is hence proportional to $E_0^2$.

Showers also create acoustic signatures \cite{1978Sci...202..749B,Askarian:1979zs}, caused by the heating and expansion of the target medium in the ionization channel around the shower axis, where the particle density is highest. This technique is studied for e.g. detection of ultra high energy neutrinos
\cite{Nahnhauer:2010uv}.

\begin{figure}[h]
\begin{center}
\includegraphics[width=9cm]{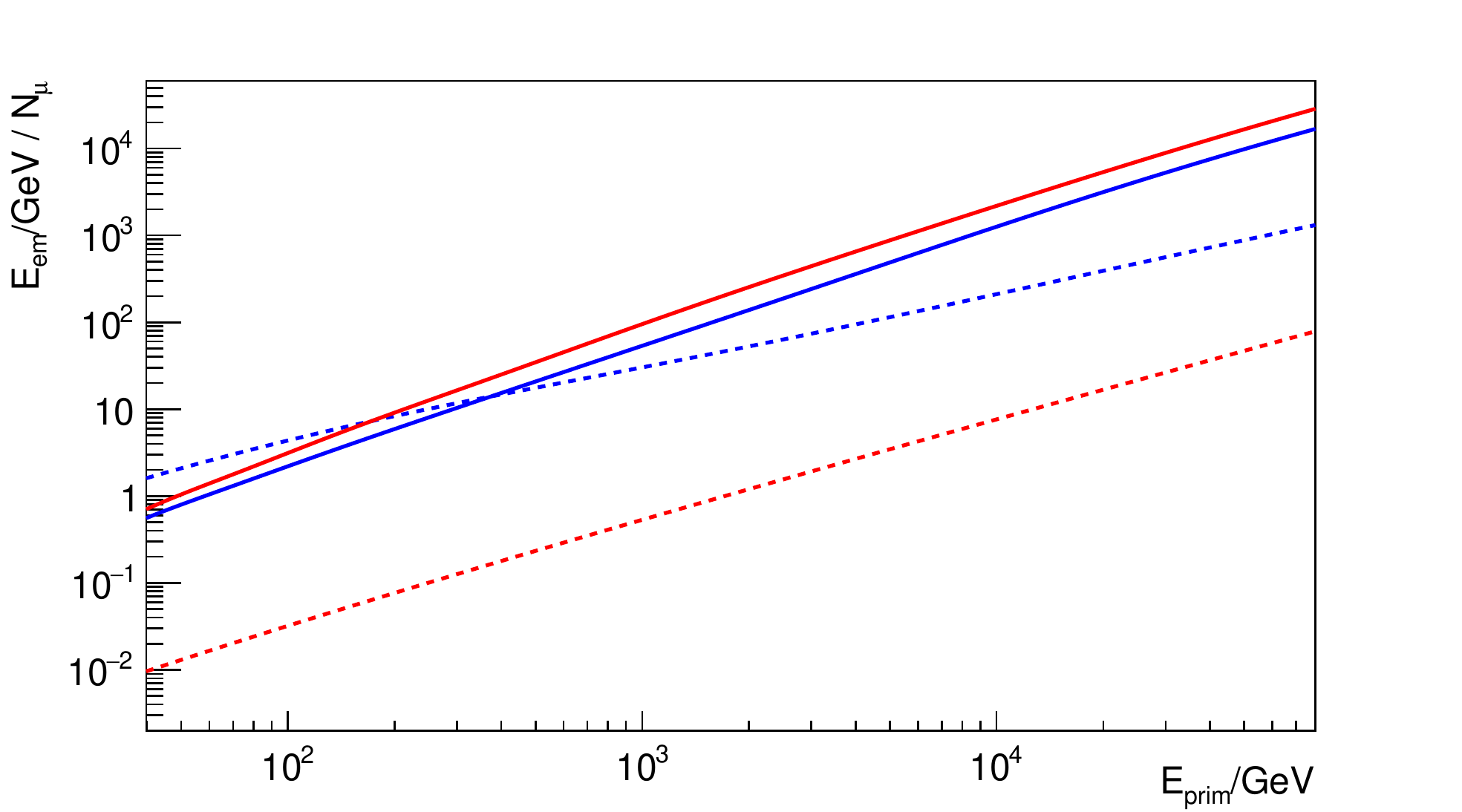}
\includegraphics[width=9cm]{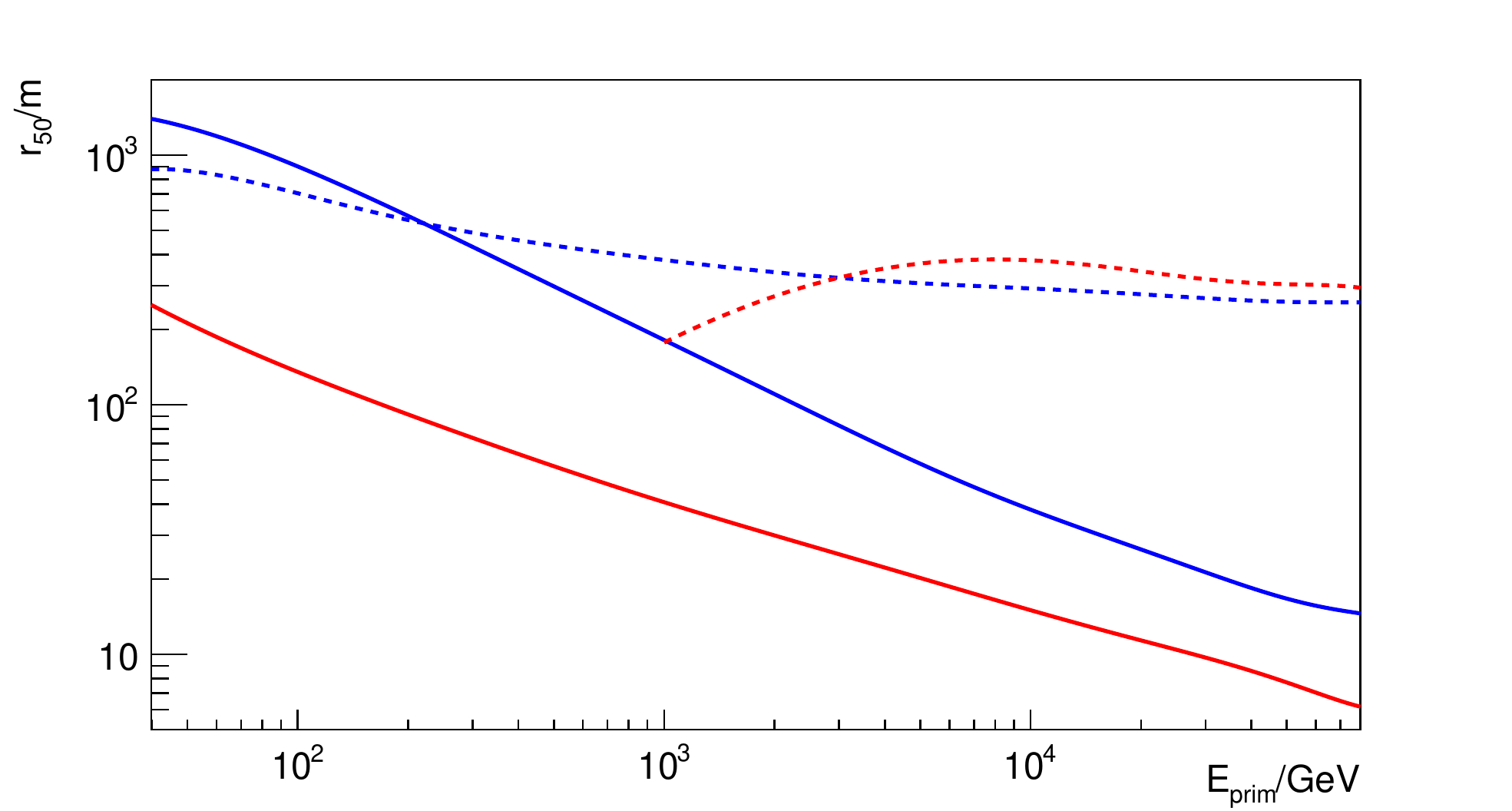}
\caption{Characteristics of air showers: particles arriving at 5000 m asl, for vertical incidence. Top,
full lines: Average energy in $e^\pm$ and $\gamma$s arriving at ground versus primary energy for gamma ray (red) and proton (blue) primaries, and (dashed lines) average number of muons arriving at ground. Bottom, full lines: radius containing 50\% of the $e^\pm, \gamma$-energy deposition for gamma ray (red) and proton (blue) primaries, and (dashed lines) radius containing 50\% of the muons (circle optimized to contain the maximum energy / number of muons, not centered on the shower axis). Based on CORSIKA simulations.}
\label{fig_sho2}
\end{center}
\end{figure}

For air showers, an alternative to these calorimetric methods is  to sample the shower particles only at ground level (at most primary energies well beyond the shower maximum, given that the atmosphere amounts to about 28 radiation lengths and 11 interaction lengths
for vertical incidence, see Fig. \ref{fig_sho1}).
The properties of particles arriving at ground level are illustrated in Fig. \ref{fig_sho2} (here the detection level is assumed as 5000 m asl, reflecting the tendency of low-threshold air shower arrays to move to high altitude). Due to $X_{\rm max}$ increasing with energy, the number of particles and the amount of energy arriving at ground increases faster than $E_0$, roughly as $E_0^{1.3}$. Gamma ray primaries and proton primaries can be distinguished by the number of muons, that is one to two orders of magnitude lower for gamma rays (with occasional muons produced via photo-production on nuclei). The radius containing 50 \% of the (electromagnetic) energy on ground decreases from scales of few 100 m at $10^{11}$ eV primary energy to 10 m scales at $10^{14}$ eV, again reflecting the deeper penetration of showers with increasing energy. Muons are scattered over scales of few 100 m. 

Shower development is a stochastic process. The energy and angular resolution for shower detectors and the capability of identifying primaries are ultimately limited by shower fluctuations.
The dominant fluctuation in shower development is the variation in the depth of the first interaction, which fluctuates by one radiation length rms ($\approx 37$ g/cm$^2$) in case of gamma ray (or electron) primaries, and by one interaction length ($\approx 90$ g/cm$^2$) in case of proton-induced showers. For hadronic showers, additional fluctuations are introduced by variations of the fraction of energy fed into electromagnetic sub-showers, mostly via $\pi^0$ production, relative to the energy going into hadrons and feeding the hadronic cascade.  Fig. \ref{fig_sho1} illustrates the shower-to-shower fluctuations, that are largest for proton-induced showers, somewhat smaller for gamma-induced showers, and modest for iron-induced showers that represent a superposition of many nucleonic sub-showers, reducing fluctuations. Given that particle density in the tail of air showers decreases exponentially on a scale of about 80  g/cm$^2$, the variation in the depth of first interaction translates into a large variation of particle number on the ground, limiting the achievable energy resolution. For air showers, the variation in the depth of first interaction also, e.g., translates into a variation of Cherenkov light yield at a level of $\Delta I/I \approx 0.3$ for gamma rays \cite{deNaurois:2009ud}, due to the height-dependent index of refraction.

\subsection{Shower Modelling and Uncertainties}

Calibration of large-volume detectors suffers from the fact that they cannot be placed in a test beam. Calibration hence depends on modeling the showers, the propagation of shower induced signals to the active detection elements, and the response of the detection elements.

For the modeling of air showers, including their Cheren-kov and radio emission, the CORSIKA code \cite{Heck:1998vt} and derivatives 
(e.g. \cite{Bernlohr:2008kv}) are standard, but other generators such as KASCADE \cite{Kertzman:1994zj} are still in use. Simulation of electromagnetic cascades uses the exact cross-section for electromagnetic processes (e.g. \cite{Rossi:1941zza,1970RvMP...42..237B}). For showers in the TeV domain, particle numbers are sufficiently small such that all shower particles can be tracked down to the lowest energies relevant for detection. Systematic uncertainties are mostly related to approximations and the finite step size in tracking \cite{Bernlohr:2008kv}. Comparing e.g. the Cherenkov light yield predicted by CORSIKA and KASKADE under otherwise identical conditions, a 5\% difference is seen  \cite{Bernlohr:2012we}.
For precise modeling of the distribution of Cherenkov light and of radio power, the exact orientation of the shower relative to the geomagnetic field and hence the elliptical spreading of the shower proved rather critical \cite{Homola:2014sra,Huege:2016veh}. 

Simulation of hadronic cascades relies on empirical\newline
hadronic interaction models tuned to reproduce accelerator data, but results are somewhat model-dependent, in particular at energies beyond current accelerator energies \cite{Engel:2011zzb}. Particle production in hadron-nucleus interactions is described by a variety of phenomenological models. Air shower modeling is very sensitive to particle spectra in the very forward direction, since these particles carry the bulk of the energy and serve to maintain the shower. This very forward region tends to be poorly covered in particular in collider particle physics experiments. Another uncertainty concerns the total proton-nucleus interaction cross section at very high energy, determining the starting height of air showers. Finally, at ultrahigh energies, tracking of all shower particles is too time-consuming, and `thinning' has to be applied, using e.g. analytical representations for sub-showers. While convergence has improved with LHC data, generators still differ by as much as 20~g/cm$^2$ in their prediction for the height of shower maximum \cite{Aab:2014kda}. Significant discrepancies, both between interaction models and between simulations and data, are seen in the predicted muon yield \cite{Aab:2016hkv,Apel:2017thr}.

\subsection{Air Shower Detection}

\begin{figure}[h!]
\begin{center}
\includegraphics[width=9.2cm]{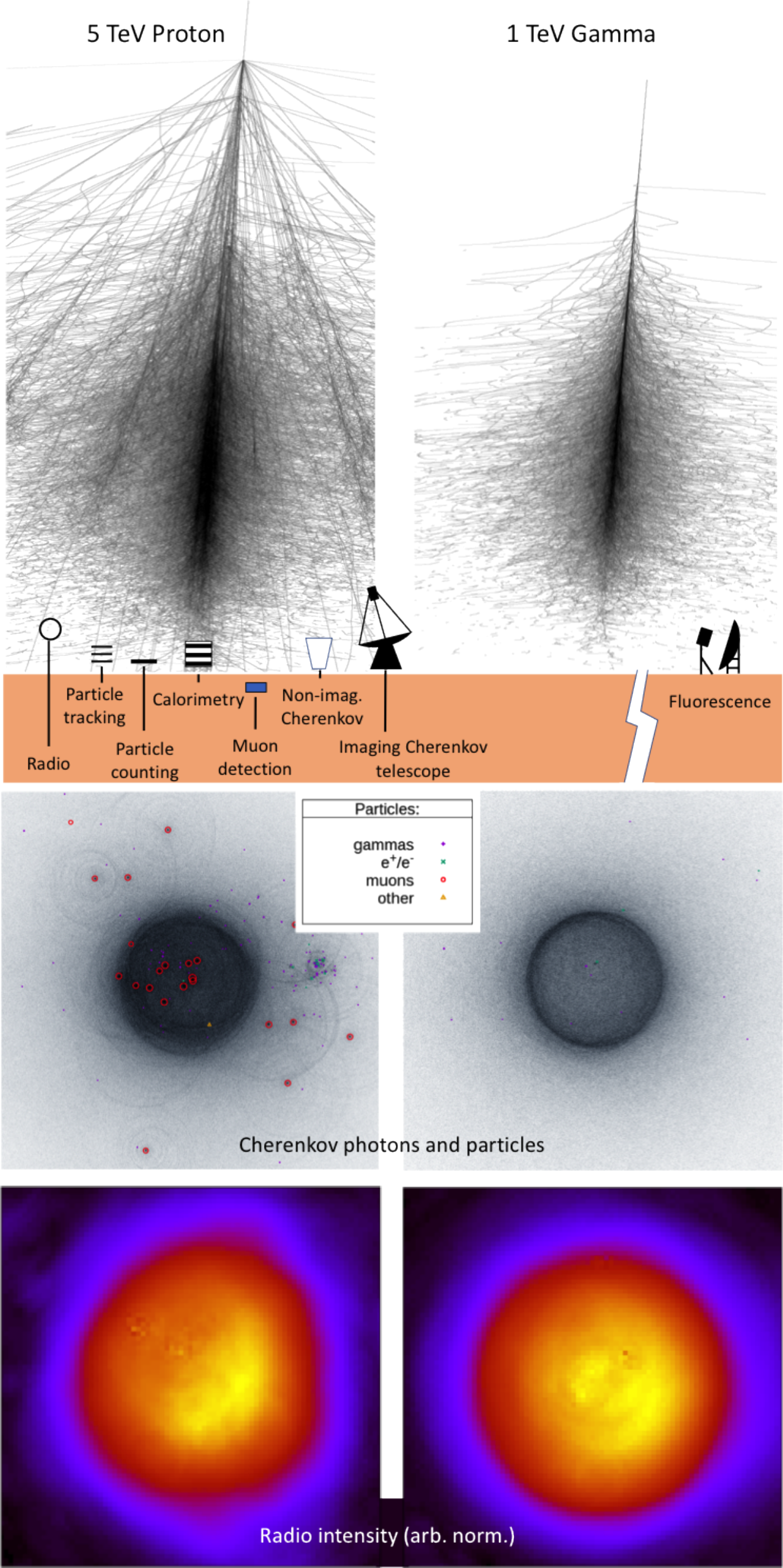}
\caption{Illustration of methods for detecting air showers. Also shown (left) is a 5 TeV proton-induced shower evolving in the atmosphere, 
with distribution of Cherenkov light (black) and particles (colored symbols) reaching the ground, and the 30-80~MHz radio intensity on the ground. 
The right panels show the same quantities for a 1~TeV gamma ray shower. For the shower displays the vertical range is 30~km, the transverse range $\pm400$\,m.}
\label{fig_asdet}
\end{center}
\end{figure}

Air shower detection and reconstruction of energy, direction and type of the primary either uses the atmosphere as an active calorimeter -- via Cherenkov, fluorescence or radio emission -- or samples the shower particles only at ground level. Alternative techniques such as the active probing of shower ionization channels using radar were proposed \cite{Gorham:2000da} but so far not demonstrated. The various detection principles are illustrated in Fig. \ref{fig_asdet}. Detection of Cherenkov light is done either in imaging mode, in which case the shower geometry is reconstructed from the Cherenkov image, often by stereoscopically observing the shower with multiple imaging telescopes from different viewing directions, or non-imaging, in which case shower direction is determined from the arrival time of the Cherenkov wavefront at different detector stations. Fluorescence light is generally viewed with (multiple) imaging telescopes. For ground-level particle detection, the shower direction is usually reconstructed from timing measurements. 
Also, tracking of shower particles has been investigated (e.g. \cite{Heintze:1988ad,Bernlohr:1995yu}), exploiting the fact that shower particles are narrowly collimated around the shower direction.

A range of techniques is applied to identify the primary particle; these rely on the measurement of the depth of shower maximum $X_{\rm max}$, (c.f. Fig.~\ref{fig_sho1}) and the width of the shower (for the shower imaging techniques), on the distribution of shower particles on the ground, and on the number of muons relative to the number of $e^\pm$ at ground level (see Fig. \ref{fig_sho2}). Proton-induced showers (see Fig.~\ref{fig_asdet}) usually contain multiple gamma-ray sub-showers, resulting in wider showers, stronger fluctuations in shower development, and a structured distribution of particles and Cherenkov light on the ground. Detection of direct Cherenkov radiation by the primary particle (enhanced by a factor $Z^2$) has been used to identify heavy primaries such as iron nuclei \cite{Aharonian:2007zja}.

\section{Ground-particle Air Shower Detectors}

\begin{figure}[h]
  \begin{center}
    \hspace{-2mm}\vspace{-1mm}
\includegraphics[width=9.7cm]{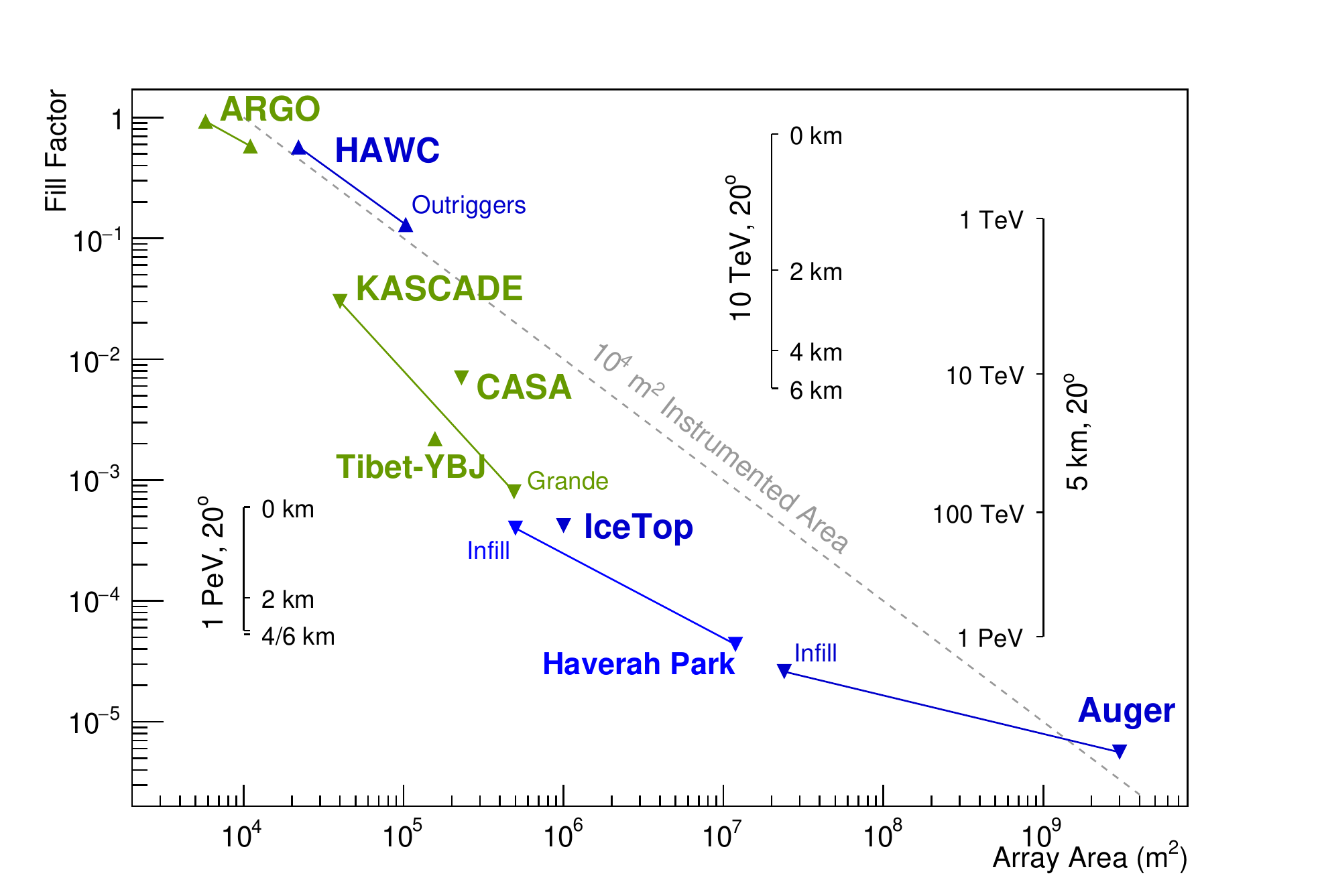}
\caption{Instrumented area fraction (fill-factor) versus total ground coverage for a selection of air shower arrays. Arrays using water as primary detection medium are shown in blue. Instruments deployed at elevations above 4000~m are marked with upward pointed triangles. All other arrays are located below 3000~m. Also shown are approximate gamma-ray threshold energies at different fill-factors and altitudes under cascade approximation B~\cite{Rossi:1941zza}.
  }
\label{fig_fillfactor}
\end{center}
\end{figure}

The most well established approach to detection of VHE and UHE
particles arriving at the Earth is via detection of air-shower
particles at ground level, recognised from the continuous background of
individual particles by a narrow coincidence window. This technique
was pioneered in the 1930s by Rossi, Auger and others and has been
used in the implementation of increasingly large and sophisticated
air-shower arrays in the decades since. For measurements in the PeV
range and above detector arrays can be rather sparse, with separation
of square metre scale detectors by hundreds of metres. Access to
lower energies requires a larger ground coverage factor and/or higher
altitude to bring the detector closer to shower
maximum. Fig.~\ref{fig_fillfactor} illustrates selected recent and some
historical arrays in terms of coverage factor and contained area. To
trigger on an air-shower and make a meaningful estimate of its
direction of origin and primary energy requires a minimum number of
detected particles and/or deposited energy within detection units
(depending on the detection approach, see
Sec.~\ref{ss_detectionunit}). Fig.~\ref{fig_fillfactor} also provides a very rough estimate of the energy range accessible to a ground-particle detector array, assuming a minimum detected shower energy of at least 50~GeV and electromagnetic showers. The appropriate value is somewhat higher for hadronic primaries as discussed in Sec.~\ref{sec_showers}. The key role played by detector altitude is evident and the need for very high fill factor (or `carpet') detectors to access the TeV range.

Fig.~\ref{fig_fillfactor} greatly underestimates the threshold for giant UHECR detectors such as the Pierre Auger Observatory~\cite{2008NIMPA.586..409A} - as these detectors are not optimised for threshold but rather for precise (and cost-effective) measurement of the highest energy particles, resulting in very large detector spacing.
The upper limit of the accessible range is determined by the array footprint and the flux of the relevant species, see Fig.~\ref{fig_cr}.

The arrival time of particles at the ground form the basis of primary direction reconstruction, with the simplest possible assumption of a plane traveling at $c$ typically extended to account for the slight curvature of the shower front. The total electron size at ground provides only a very rough indication of primary energy due to fluctuations in the depth of shower maximum~(see Sec.~\ref{sec_showers}). This is one of the main motivations for `hybrid' detectors, 
which make simultaneous observations with ground particle and shower imaging detectors, capturing the shower particle content and depth of maximum together. The most prominent of the hybrid detectors is the Pierre Auger Observatory and the hybrid approach is illustrated for Auger in Sec.~\ref{sec_imaging}.

\subsection{Considerations in the design of a detection unit}
\label{ss_detectionunit}

The basic objectives for an individual element in a ground-particle
air-shower detector are the measurement of the local shower particle
density and arrival time. Beyond this some experiments allow
separation of particle types (in particular muon separation from
electrons and gammas) and may allow the time structure of the shower
front to be extracted.

For the measurement of the electromagnetic cascade contents the primary choice is between particle counting and calorimetery. A thin ($\ll$1 radiation length) detector provides a signal proportional to the number of charged particles, which is dominated by electrons except at very large zenith angles (see Sec.~\ref{sec_showers}). The addition of a high-$Z$ conversion layer (often a sheet of lead of thickness $\sim$5~mm $\sim$ 1$\lambda_{r}$) increases the particle count through pair-production of the more numerous gamma-ray component. A common solution for thin detectors is a plastic scintillator coupled to a photomultiplier tube or other photosensor via an (often wavelength shifting) light guide. Resistive plate chambers (RPCs) have also been employed in air shower arrays and may offer advantages in terms of time resolution in return for the additional operational complexity due to the need for gas replenishment (see e.g.~\cite{2015APh....67...47B}).

A thick (many radiation length) detector enables electromagnetic cascade development and hence extraction of a signal proportional to the total incident energy (for the electromagnetic component). As the average energy of particles falls with distance from the core the lateral distribution is somewhat steeper for a calorimetric detector. HAWC is an example of a high-altitude, high fill-factor array based on many radiation lengths of water and focused on $\gamma$-ray astronomy~\cite{2012NIMPA.692...72D}.

For the purposes of cosmic ray mass composition sensitivity, or background rejection for gamma-ray astronomy, it is desirable to separate and count individual muons in a detector unit (see Fig.~\ref{fig_sho2}).
Such identification requires many radiation lengths of material, sufficient for most of the cascades initiated by electrons or photons of up to at least GeV energies to have died out: $X/\lambda_r >  \sim 10$ but still introducing ionisation energy losses
much smaller than the typical energy of muons in showers of a few GeV. 
Low cost solutions to provide the cascade attenuation include water, which can be used as calorimeteric EM detector as well as absorber, and rock/earth. 

Apart from area coverage and depth the critical performance criteria for a detector unit are its time and amplitude resolution. The time resolution impacts directly on the angular resolution if it is not small in comparison with the thickness of the shower front times $c$. The arrival time distribution of particles has a long tail, particularly at large impact distance, but the rising edge has typical timescales of $\sim$1~ns. The larger the baseline of the detector the more relaxed the timing requirements become to maintain the same resolution: an error of $\Delta t$ over a baseline $b$ corresponds to a $\sim0.2 (\Delta t / 1 \mathrm{ns})/(b / 100 \mathrm{m})$ degree directional error.

As fluctuations in particle number are typically rather large except for very large area detectors and/or high primary energies and/or small impact distances, the required amplitude resolution is modest ($\sim$30\%),
but a large dynamic range is highly desirable to deal with the strong impact distance dependence of the signal.
It may often be impractical to optimise for both time and amplitude resolution, for example for a water-tank based detector the choice of a reflecting inner surface provides a closer to calorimetric signal, but absorbing walls result in a narrower arrival time distribution. 

Note that many more sophisticated unit detectors have been employed over the years includes those with the capacity to track individual particles
and provide detailed measurements of shower-core hadrons
but such detectors have been most successful in testing interaction and shower models rather than in general purpose cosmic ray or gamma-ray astronomy. KASCADE is arguably the best example of a multi-technology ground particle array operating at intermediate (around PeV) energies~\cite{2003NIMPA.513..490A}.

\subsection{Array Performance}

The primary performance criteria for an air-shower array are effective area, threshold, angular resolution and separation power / composition sensitivity.
The detection area of a ground particle detector is essentially the array area for shower energies well above threshold. For arrays with a linear scale that is not much larger than the typical shower footprint (see Fig.~\ref{fig_sho2}), some kind of guard-ring or `outrigger' detectors are sometimes employed to make sure events landing at the detector edge can still be properly reconstructed.

The threshold energy of an array is dictated primarily by the combination of altitude and fill-factor, as illustrated in Fig.~\ref{fig_fillfactor}. Beyond this, the threshold of an individual detector unit and the trigger (or event identification/selection) at the array level play a role, with accidental coincidences as the limitation for array level triggering. The rate of individual muons is $\sim$200~Hz per m$^{2}$ of detector, with modest dependence on altitude, necessitating large coincidence multiplicities to avoid very high accidental trigger rates in a large array. The threshold energy is strongly zenith angle dependent and hence the rate of detected showers falls off rapidly. Close the horizontal (beyond typically 70$^{\circ}$) the electromagnetic component is negligible and muons are separated by the geomagnetic field in to $\mu^{+}$ and $\mu^{-}$ with radial symmetry lost. The occurrence of a symmetric or EM-rich shower at large zenith angle is therefore a signature of a neutrino primary penetrating deep in to the atmosphere before interacting.

The angular resolution of an array is dictated by the quality of the individual time measurements and the number and baseline of such measurements as discussed above. Resolutions of around 1$^{\circ}$ are usually achievable without extreme effort, 0.1$^{\circ}$ represents the current limit for the technique. 
The overall sensitivity of a detector of this type can be estimated following the approach given in Sec.~\ref{sec_det}. The cosmic ray shadows of the moon and the sun, as well as the bright, steady and very close to point-like signal from the Crab Nebula, provide a useful means of confirming resolution and energy reconstruction / flux derivation. 

Ground level observables that provide information on the primary particle type can be used to do cosmic ray astrophysics and are used as background rejection technique for gamma-ray astronomy. The primary distinguishing features are the shape of the EM lateral distribution and the number of muons in the shower relative to the EM size, Fig.~\ref{fig_sho2} gives an indication of the detector requirements for effective separation using muons. In addition, time domain information, including shower front curvature and signal rise-time, provide information on shower development and hence on the primary nature.

The challenge for neutrino detection using such arrays is the limited solid angle within which incoming neutrinos can be detected and identified. Only for close to horizontal events is air-shower development and unambiguous identification possible. For such showers the detector height becomes critical rather than the top surface and the array is dramatically foreshortened in the shower plane. For showers above the horizontal the target material for neutrino interaction is the up to $\sim$40~kg/cm$^{2}$ of air, for slightly upward going events, so-called Earth-skimming neutrinos, the target is rock or ice, with the depth set by the energy losses of the neutrino interaction product. The most promising Earth-skimming case is for tau-neutrinos, as at ultra-high energies the tau can travel through kilometers of rock and generate a cascade in the air if it decays after emergence.


\section{Imaging of Air Showers}
\label{sec_imaging}

Imaging of the shower trajectory with telescopes, either using Cherenkov emission beamed along the shower axis, or the isotropic fluorescence emission, has proven a very successful approach towards determing shower geometry and hence the direction of the primary, its energy (from the image intensity), and its type (from the longitudinal and lateral image profile). Cherenkov emission with its concentrated light pool allows low energy threshold -- in the 10 GeV range for gamma rays and imaging telescopes in the 20 m diameter range -- but limits effective area per telescope to $O(10^5)$ m$^2$, whereas fluorescense emission -- where viewing distance is primarily limited by atmospheric extinction -- can provide huge effective areas in the $10^8$ m$^2$ range per wide-angle telescope, but coupled with high energy thresholds, typically around $10^{17}$ eV for telescope apertures in the few-m range, as currently used. Due to the beamed nature of the Cherenkov emission, Cherenkov telescopes need to point at the source, providing a field of view for primary particles roughly equal to the optical field of view of the telescope, whereas fluorescence telescope can view arbitrary shower directions, and therefore usually have a fixed, near-horizontal pointing. Virtually all modern systems use multiple telescopes to provide stereoscopic views of the shower trajectory, for unambiguous determination of the direction of the primary.

\subsection{Cherenkov Imaging}

Stereoscopic arrays of Imaging Atmospheric Cherenkov Telescopes (IACTs) have emerged as the most powerful narrow-angle air shower detectors, in particular for very high energy gamma ray astronomy \cite{Weekes:2005da,Lorenz:2012nw,deNaurois:2015oda}. In IACT arrays, the air shower is reconstructed from several (at least 2) different views, providing a purely geometrical reconstruction of the shower axis in space and of the impact point (e.g. \cite{Hofmann:1999rh}). 
Current IACT systems such H.E.S.S. \cite{Aharonian:2006pe}, MAGIC \cite{Aleksic:2014lkm} and VERITAS \cite{Park:2015ysa} comprise 2 to 5 telescopes with 100 to 600 m$^2$ mirror area; the spacing of telescopes is around 100 m, as a compromise between the large baselines desired for good stereoscopic imaging, and the requirement that telescope spacing should be well below the diameter of the Cherenkov light pool, so that telescopes are illuminated simultaneously. In such geometry, many showers that trigger two or more telescopes impact well outside the footprint of the telescope system, resulting in near-parallel views and degraded reconstruction of shower geometry. The next-generation Cherenkov Telescope Array \cite{Acharya:2013sxa,Acharya:2017ttl} with 99 telescopes in its southern array for the first time provides an array with a footprint large compared to the size of the light pool, ensuring that (a) the light pool is contained for a significant fraction of events and hence a larger number of images is provided for each shower, covering the full range of viewing angles, and (b) showers are efficiently triggered since for contained showers there are always telescopes in the region of highest Cherenkov light intensity. 

\begin{figure}[h]
\begin{center}
\vspace{3mm}
\includegraphics[width=0.45\textwidth]{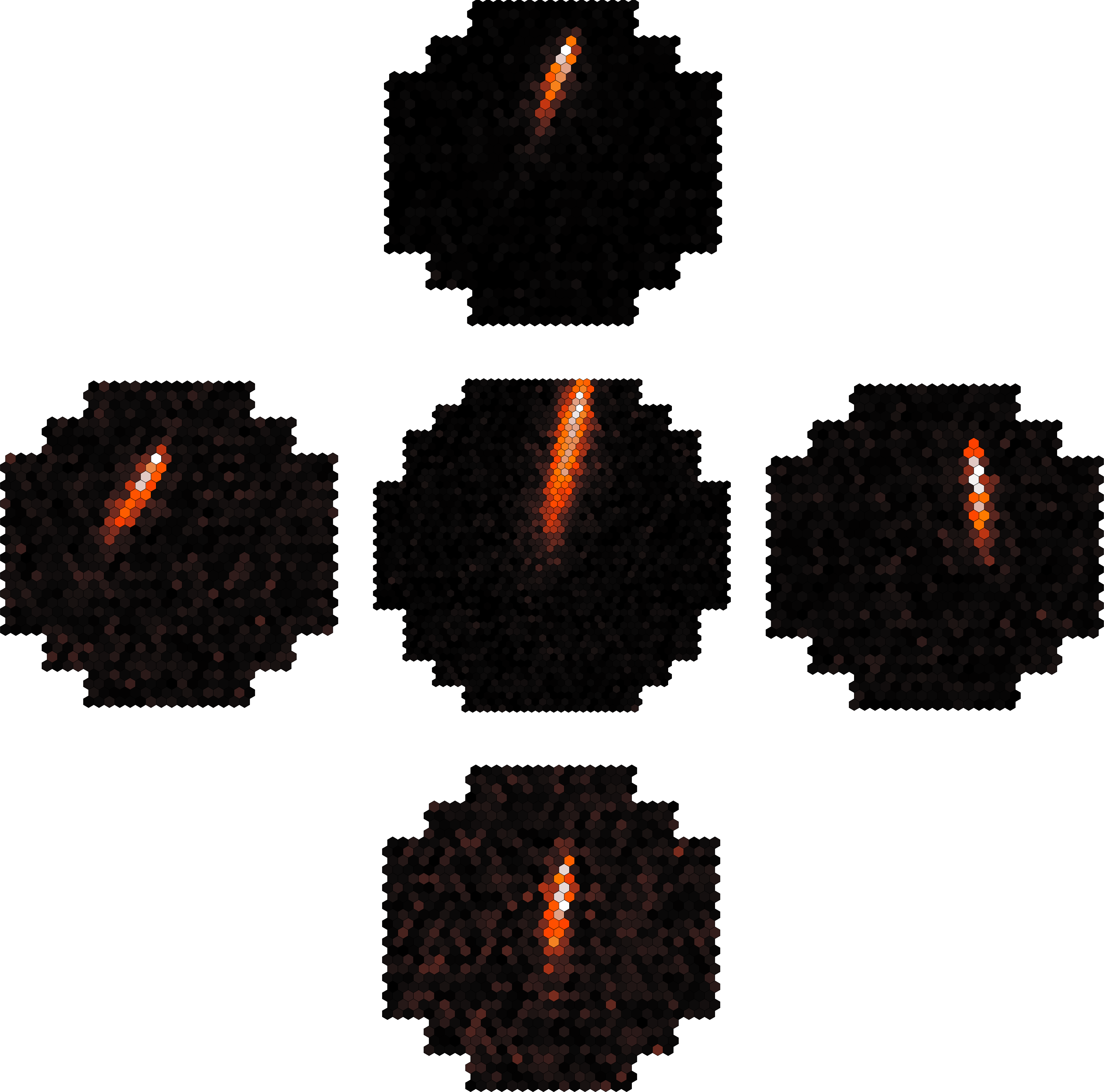}\\
\vspace{0.5cm}
\includegraphics[width=0.48\textwidth]{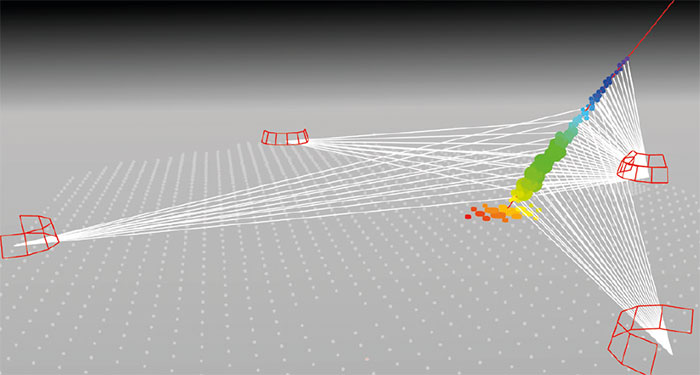}
\caption{Top: Images of an air shower viewed with the five H.E.S.S. IACTs, from different viewing angles. The central telescope hat larger mirror area and hence a more intense image. Credit: H.E.S.S. Collaboration.
Bottom: Event display of an ultra high energy air shower detected with Auger, showing surface detectors recording hits and light seen along the air shower track by the four fluorescence-detector sites. Circle size corresponds to intensity, and color to time. Credit: Pierre Auger Collaboration.}
\label{fig_images}
\end{center}
\end{figure}

Viewed from an impact distance $d$, the typical energy deposition of a TeV gamma-ray air shower, peaking at about 350~g/cm$^2$ (8~km asl) with a width of $\pm 150$ g/cm$^2$ (6 to 12 km asl) ((Fig. \ref{fig_sho1}), results in an elliptical image of 
about $0.3^\circ (d/100\,\mbox{m})$ rms length, and $0.1^\circ$ to $0.2^\circ$ width (Fig. \ref{fig_images} top). 
Images are analyzed either by parametrizing images in terms of their orientation, width and length (`Hillas parameters'), followed by geometrical reconstruction and selection based these image parameters, or by fitting analytical \cite{deNaurois:2009ud} or numerical models \cite{Parsons:2014voa} of the shower image to the data. For gamma-ray/cosmic-ray separation and also for shower reconstruction, machine learning techniques are increasingly employed.

The energy threshold of Cherenkov telescopes is governed by two factors, (a) a minimum of 30 to 50 detected photons are required to define a shower image; for a density of detected photons of about $10$\,m$^{-2}$TeV$^{-1}$ this implies a threshold for gamma rays of $\approx 5$ TeV/$A_{\rm m^2}$ where $A$ is the mirror area. And (b) that this signal must stand out above the night sky photons that amount to about 40\,MHz per $m^2$ mirror area and square degree in the focal plane. Requiring the signal in a typical $0.25$ square degree image region to be well above the night sky noise implies for a 5\,ns signal integration time a threshold of $E_{\rm TeV} > 0.2 / A^{1/2}_{\rm m^2}$. The comparison of the two terms shows that up to mirror areas approaching 1000\,m$^2$, the first term dominates. Since the density of Cherenkov light decreases with the square of the distance to the shower maximum, thresholds can be lowered  by a factor $\approx 2$ by placing telescopes at 5 km asl rather than the typical 2 km asl \cite{Aharonian:2000rf}. 

\begin{figure}[h]
\begin{center}
\includegraphics[width=0.48\textwidth, trim=90 32 80 18, clip]{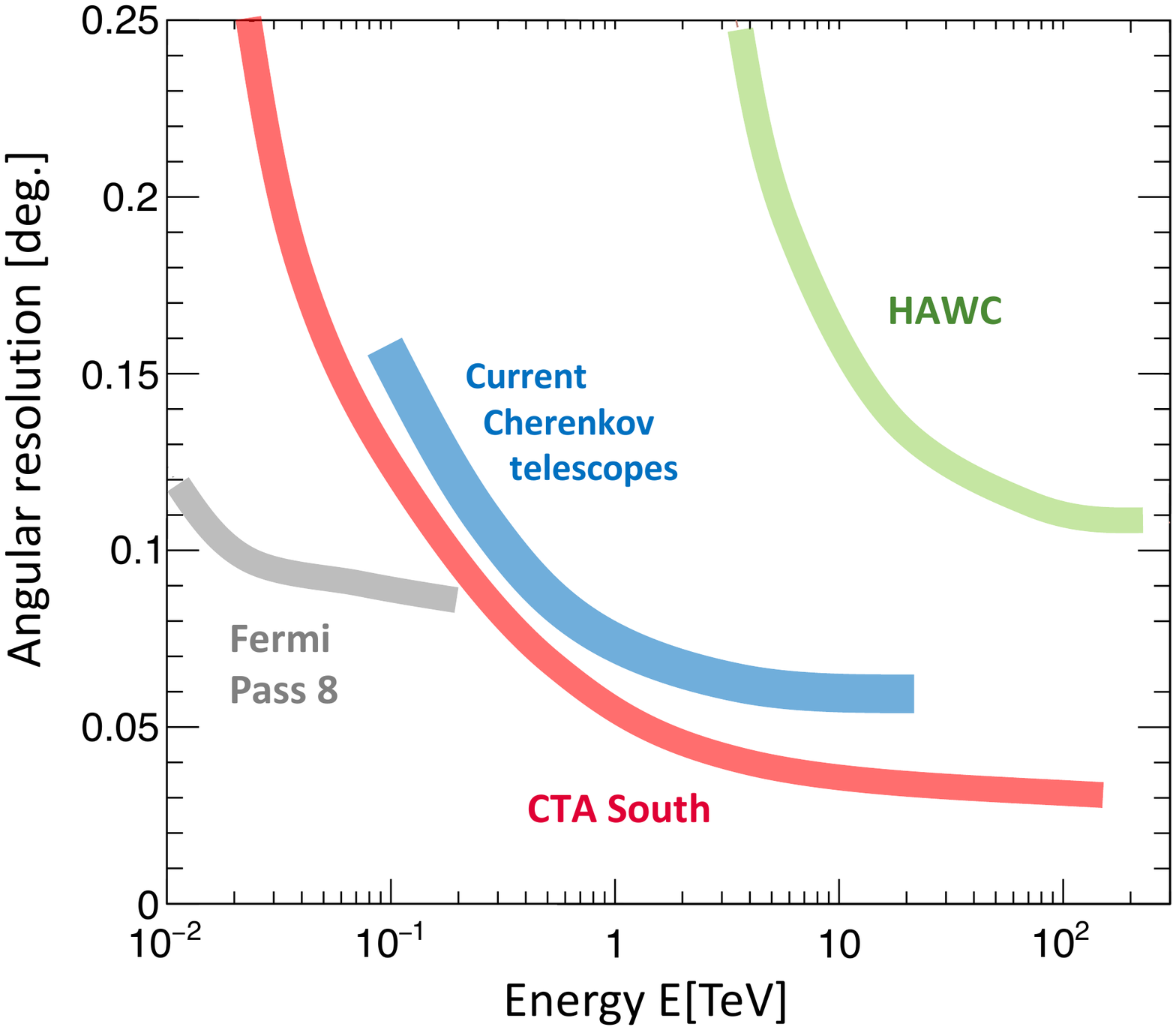} 
\caption{Angular resolution of gamma-ray detectors: Fermi-LAT (Pass 8), current Cherenkov telescopes (HESS, MAGIC, VERITAS), and HAWC.  Adapted from \cite{Acharya:2017ttl}}
\label{fig_angres}
\end{center}
\end{figure}

Not too close to their energy threshold, current IACT systems reach a typical 68\% containment angular resolution $\theta_{68}$ of about 5 arc-min, see Fig. \ref{fig_angres}. For gamma ray energies above few 10s of GeV, the angular resolution is mostly determined by photon statistics in the images (and somewhat by camera pixel size), and can be improved by collecting more photons per telescope, or more images per shower; for larger numbers of telescopes, resolutions improve like $1/N_{\rm image}^{1/2}$. The physical resolution limit of the IACT technique due to shower fluctuations has been estimated to $0.4'/ E_{\rm TeV}^{1/2}$ \cite{Hofmann:2006wf}.

\begin{figure}[h]
\begin{center}
\includegraphics[width=0.48\textwidth, trim=40 34 42 18, clip]
{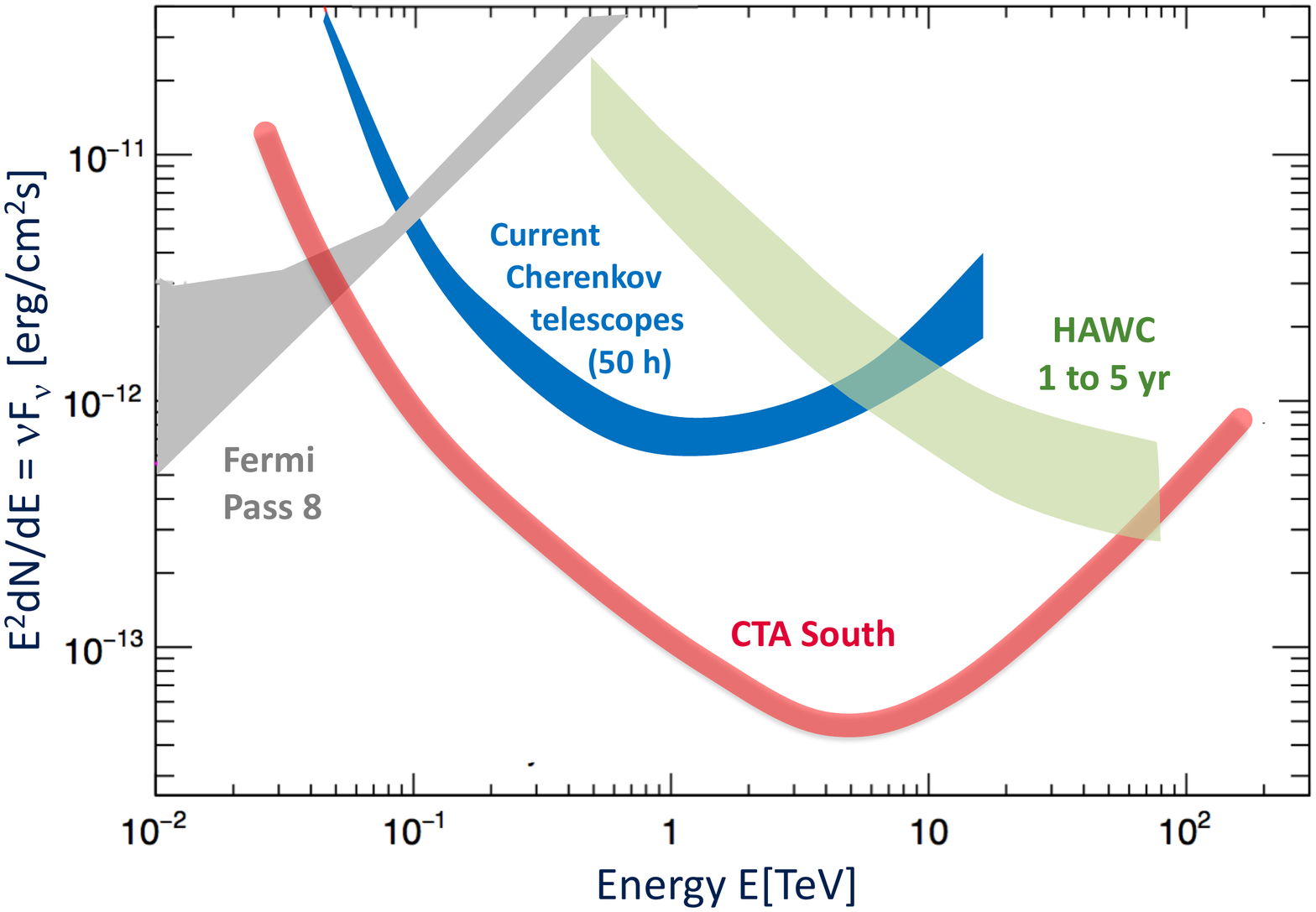}
\caption{Sensitivity of gamma-ray detectors: Fermi-LAT (Pass 8, $\approx 5$ years); current Cherenkov telescopes (HESS, MAGIC, VERITAS, 50~h exposure); HAWC (1 to 5 years exposure). The sensitivity corresponds to five standard deviation detections in five independent logarithmic bins per decade in energy. For the IACT sensitivities, additional criteria are applied to require at least ten detected gamma rays per energy bin. Adapted from \cite{Acharya:2017ttl}}
\label{fig_sens}
\end{center}
\end{figure}

The differential flux sensitivity for gamma ray detection is frequently defined as a $5 \sigma$ detection per $\Delta \log_{10}{E} = 0.2$ interval (5 bins per decade), with at least $N_{\rm min} = 10$
detected gamma rays. Sensitivity is quoted as $E^2 \phi_{\rm min} = \left( \nu f_\nu \right)_{\rm min}$, in units of minimal detectable energy flux, erg/ cm$^2$s.
Fig. \ref{fig_sens} illustrates sensitivity for some of the current and planned gamma-ray instruments.
At the highest energies, sensitivity is governed by the requirement to detect at least $N_{min}$ gamma rays, resulting in 
$E^2 \phi_{\rm min} \approx 2 N_{\rm min} E (A_{\rm eff} T)^{-1}$
$\approx 10^{-6} E_{\rm TeV}$ $(A_{\rm eff,m^2} T_h)^{-1}$ erg/cm$^2$s, where $T$ is the observation time and $A_{eff}$ includes all selection cuts applied. The energy flux sensitivity increases linearly with energy, reflecting the energy carried by each detected gamma ray.
At intermediate energies in the TeV range, modern IACT systems provide sufficient rejection of cosmic ray nucleons that sensitivity is largely governed by the electron background flux $\phi_e$ (see Fig. \ref{fig_cr}). While it is possible to suppress electrons, e.g. by cuts on the height of the shower maximum, this usually causes significant loss in gamma ray efficiency, so that effective sensitivity is not improved. The sensitivity is then 
$E^2 \phi_{\rm min} \approx 10 E^2 \phi_e (A_{\rm eff} T  E \pi \theta^2)^{-1/2}$, where $\theta$ is the angular cut applied to define the source region, either the source size of extended sources, or $\theta \approx 1.5 \theta_{68}$ for point sources. 
With
$\phi_e(E) \approx 1.2 \cdot 10^{-4} E^{-3}_{\rm TeV}$ m$^{-2}$s$^{-1}$sr$^{-1}$TeV$^{-1}$ for the range of tens of GeV to 1 TeV, one finds
$E^2 \phi_{\rm min} \approx 2 \cdot 10^{-10} \theta_{68,'}$  $(A_{\rm eff,m^2} T_h)^{-1/2}$ erg/cm$^2$s, where $\theta_{68,'}$ is in arc-min. Energy flux sensitivity
in this energy range is approximately constant. At low energies, near the IACT threshold, cosmic ray rejection and angular resolution and hence sensitivity rapidly deteriorate. 

The primary application of Cherenkov telescopes is very high energy gamma ray astronomy. Since the bulk of air showers detected by Cherenkov telescopes are showers induced by charged cosmic rays, telescopes are also used to study these cosmic rays. Applications include the measurement of proton energy spectra at TeV energies (e.g. \cite{Aharonian:1999zw}) and the measurement of the multi-TeV flux of iron nuclei (e.g. \cite{Aharonian:2007zja}) that are identified via the direct Cherenkov light that the primary nucleus emits high up in the atmosphere, and that is detectable since its intensity in increased by a factor $Z^2$ compared to a single-charged particle. Probably the most important non-gamma-ray application is the measurement of the cosmic-ray electron/positron flux in the TeV range, identified as a uniform flux of particles initiating electromagnetic showers, whose average height of shower maximum indicates that the initiating particles are electrons rather than diffuse gamma rays (e.g. \cite{Aharonian:2009ah}).

An alternative to imaging the shower Cherenkov light onto a pixelated camera are arrays of wide-angle Cherenkov light sensors 
\cite{Karle:1995dk,Tluczykont:2014cva}. Their shower reconstruction techniques resemble those of ground particle air shower arrays; shower core location is determined from the light intensity measured across the array, shower direction from differences in light arrival time across the array. Unlike ground particle arrays, wide-angle Cherenkov arrays provide a calorimetric energy measurement, with improved energy resolution. For use in gamma ray astronomy, the main drawback is the modest separation power between gamma ray showers and cosmic ray showers; the main advantage lies in the low cost per effective area, allowing large arrays (up to 100~km$^2$ in the HiSCORE \cite{Tluczykont:2014cva} proposal) that improve sensitivity at the count-starved highest energies, with sensitivity in the $10^{-13}$ erg~cm$^{-2}$s$^{-1}$ range from about 100~TeV to 5~PeV. While existing (smaller) such arrays contribute significantly to cosmic-ray studies, their successful use for gamma ray astronomy remains to be demonstrated.

\subsection{Fluorescence Imaging}

Fluorescence imaging is so far primarily used for imaging of ultra-high-energy (UHE) air showers (Fig. \ref{fig_images} bottom). Energy deposition from charged air-shower particles leads to the excitation of nitrogen molecules in the atmosphere, which spontaneously de-excite by the emission of UV (fluorescence) light in the 290 to 430~nm range. The fluorescence yield of $\approx$7 photons per MeV ionization energy \cite{Arqueros:2008cx,Patrignani:2016xqp} provides a calorimetric measurement of the energy of the primary.
Due to the isotropic nature of the emission, for UHE air showers the fluorescence light can be detected very far from the shower axis, and at arbitrary angles to the shower axis; neglecting atmospheric absorption, the light yield of a shower of energy $E_{17}$ (in units of $10^{17}$ eV) observed at distance $d_{10}$ (in units of 10 km) with telescope with area $A_{\rm m^2}$ and photon detection efficiency $q$ is $n_{\rm photon} \approx 6 \cdot 10^2 q\,A_{\rm m^2} E_{17}/d_{10}^2$. 
Shower geometry can be reconstructed using two stereo telescopes, or using a single telescope, the time gradient of the image, and the shower impact point measured with a surface array of detectors. The calorimetric fluorescence imaging -- with a duty cycle of 10-15\% -- can be used for cross-calibrating  the surface detector, and for providing a $X_{\rm max}$ measurement for a fraction of the events, for use in composition measurements. Both the Auger \cite{Abraham:2009pm} and Telesope Array \cite{Tokuno:2012mi} detectors follow this strategy, as does the LHAASO array under construction \cite{Liu:2015yba}. Even at 10s of km distance, an air shower viewed sideways subtends an angle of few $10^\circ$ and the imager must provide a large FoV. Auger e.g. uses 4 fluorescence detector stations at the circumference of the 3000 km$^2$ surface array for stereoscopic viewing, each station equipped with 6 independent telescopes with $30^\circ \times 30^\circ$ FoV, covering $180^\circ$ in azimuth. Three HEAT telescopes cover $30^\circ$ to $60^\circ$ elevation angle, allowing close by lower-energy showers to be tracked.
The exposure of a fluorescence telescope can be crudely estimated as $A_{eff} T \approx 10^{14} ( 6 \cdot 10^2 / n_{\rm min}) (\phi/30^\circ) q\,A_{m^2} E_{17} T_{\rm yr}$ m$^2$s where $T_{\rm yr}$ is the running time in calendar years (with $\approx 10\%$ duty cycle for fluorescence operation), and $\phi$ the azimuthal acceptance of the telescope, indicating that with multiple telescopes of $q\,A_{\rm m^3} \approx 1$ and multi-year operation exposures in the range of $10^{16}$ m$^2$s are feasible, as required for UHE cosmic ray detection (see Fig. \ref{fig_cr}). The JEM-EUSO program \cite{Fenu:2017xct} aims to deploy
wide field of view fluorescence cameras in space, looking down onto the atmosphere, boosting exposure by an order of magnitude in comparison to ground-based systems, albeit with an energy threshold of a few $10^{19}$ eV.

\subsection{Instrumentation Aspects for Cherenkov and Fluorescence Imaging}

In terms of technology of mirrors and focal plane, fluorescence telescopes (FTs) and imaging atmospheric Cheren-kov telescopes (IACTs) share many common features. Main differences are that FTs are usually not moveable, have wider fields of view (FoV) (up to $30^\circ$ for FTs as opposed to $5^\circ$ to $10^\circ$ for IACTs) and usually coarser pixels to cover the wide field with an affordable number of pixels ($1^\circ$ to $2^\circ$ as opposed to $0.1^\circ$ to $0.2^\circ$).  Most instruments rely on photomultiplier tubes for light detection, usually equipped with Winston cone light concentrators to eliminate gaps between pixels and also to limit the viewing angle of a pixel; silicon sensors start to come in use in IACTs. Signals are slower for FTs -- a shower track viewed perpendicular to the track at 20\,km distance propagates across the image at 
$\approx 1 \mu$s/deg., whereas Cherenkov light and shower particles propagate almost co-linearly at the speed of light, resulting in a few-ns width of the Cherenkov pulse. Hence signals are sampled in the 10 MHz range in case of FTs, and in the GHz range for IACTs. Trigger electronics reacts to pixel coincidences on the $10 \mu$s time scale for FTs, and on the 10 ns scale for IACTs.

Telescopes mostly use single-mirror optics, with mirrors composed of facets arranged in Davies Cotton geometry (facets of fixed focal length mounted on a sphere of radius $r=f$) or paraboloid geometry (mirrors arranged on  a paraboloid $z = r^2/(4f)$).  In order to achieve sufficiently good imaging over an extended field of view, relatively large (compared e.g. to radio telescopes) ratios of focal length $f$ to diameter $d$ are required: the point spread function at field angle $\delta$ is governed by coma, $\sigma \approx 0.03\delta/(d/f)^2$ \cite{Schliesser:2005qu,Vassiliev:2006pw}. IACTs typically use $f/d$ in the 1.2 - 1.6 range, to provide sufficiently good imaging over a FoV of $5^\circ$ to $10^\circ$ diameter.

Towards providing imaging over wider FoV, as required for FTs and desirable for IACTs for surveys, study of extended sources, and follow-up of poorly-located transients, different options exist. The optics of wide FoV single mirror telescopes is discussed in \cite{Schliesser:2005qu}. Schmidt-optics with a corrector ring is used in the Auger FTs \cite{Abraham:2009pm}; a $15^\circ$ FoV Schmidt IACT with a Fresnel corrector plate was considered in \cite{Mirzoyan:2008pj}.  MACHETE \cite{Cortina:2015xra} is a proposal for dual non-steerable Schmidt-optics IACTs with 45 m $\times$ 15 m mirrors, providing $60^\circ \times 5^\circ$ FoV.
Schwarzschild-Couder dual mirror IACTs \cite{Vassiliev:2006pw} provide a FoV up to $15^\circ$ with good PSF; the 9.7 m CTA-SCT telescope prototype with 5.4 m secondary is under construction \cite{Byrum:2015rla} and the  4.3 m CTA-ASTRI telescope prototype with 1.8 m secondary, $10.5^\circ$ FoV and 2368 pixels is operational \cite{Giro:2017vuk}. 

Various techniques are used to manufacture mirror facets \cite{Forster:2013txa}, all relying on an aluminum reflective surface that provides high reflectivity in the $\sim$300\,nm to $\sim$600 nm range: in use are  polished and coated glass mirrors, precision-machined aluminum mirrors, and mirrors produced by mold-replication techniques, using a cold-slumped or (for small radii of curvature) hot-slumped glas sheet, backed e.g. by a honeycomb structure and another glass sheet. Aluminum surfaces are protected by silicon-dioxide or multi-layer coatings.

Lacking cosmic `test beams' the energy calibration of FTs and IACTs heavily relies on simulations of light generation, propagation and detection, that in turn require rather precise knowledge of atmospheric conditions, see e.g, \cite{Bernlohr:2000wq} and \cite{Abraham:2010pf} for comprehensive discussion for IACTs and FTs.
The production of Cherenkov light is known from first principles. The fluorescence yield, on the other hand, needs to be measured with laboratory beams, and may be influenced by humidity \cite{Arqueros:2008cx}; a precision of 3.5\% is reached \cite{Rosado:2014bya}.  
Measurements made with the Electron Light Source at Telescope Array confirmed the extrapolation of lab measurements to the field \cite{Shibata:2008zzg}.
Precise modeling of air showers relies on knowledge of atmospheric density profiles, that influence both the shower evolution and the yield of signals such as Cherenkov photons. Differences between summer and winter atmosphere at a mid-latitude location can e.g. cause up to 20\% difference in the yield of Cherenkov light at ground \cite{Bernlohr:2000wq}. Radiosondes and global atmospheric models are being used to obtain time-dependent profiles as input to simulations, and to reduce related uncertainties \cite{Abreu:2012zg}. Atmospheric transmission cuts off below 300~nm due to Ozone absorption; at longer wavelengths, molecular (Rayleigh) and aerosol (Mie) scattering determine the transmission; at 400~nm the vertical transmission from 10~km height to ground is about 70\%, approaching 90\% at 600~nm \cite{Bernlohr:2000wq}
Significant progress has been achieved e.g. by use of Lidar systems to characterize the back-scattering and hence the aerosol density as a function of height, under actual observing conditions \cite{Abraham:2010pf,Fruck:2014mja}. The achievable precision in the absolute energy calibration in IACTs and FTs is estimated to $\approx 15\%$ \cite{Gaug:2017nci} and $\approx 20\%$ \cite{Abraham:2010mj}, respectively.

\subsection{Radio `Imaging'}

\begin{figure}[h]
\begin{center}
\vspace{-3mm}
\includegraphics[width=9.6cm]{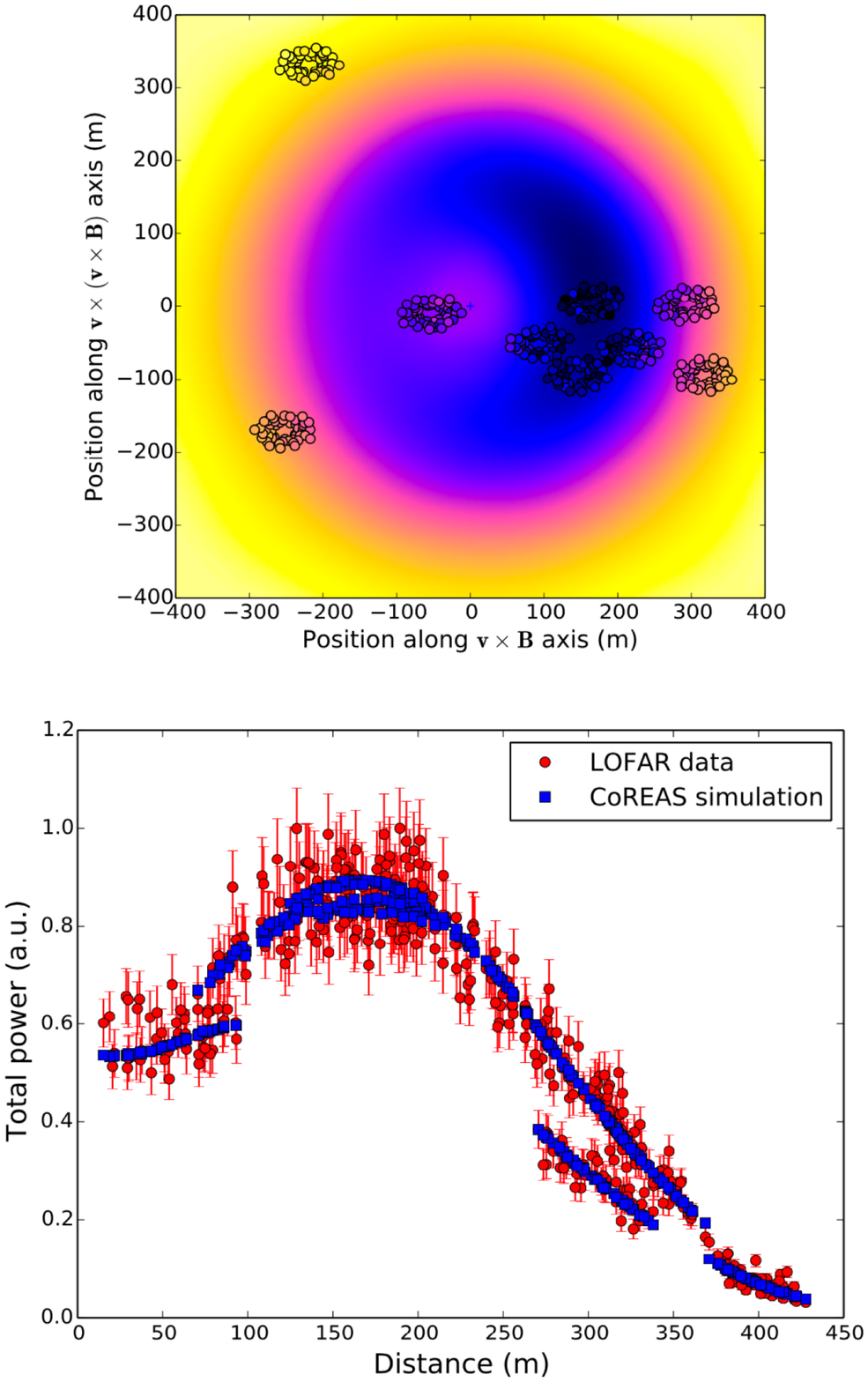}
\vspace{-7mm}
\caption{Radio power distribution and comparison with simulations. Top: The 30-80 MHz radio power measured with LOFAR stations is fitted to a CoREAS simulated radio map (top); the power distribution clearly shows the azimuthal asymmetry resulting from the interference of geomagnetic and Askaryan emission.  Bottom: Measured power (red) and simulation (blue) as a function of distance to the shower axis. Because of the azimuthal asymmetry, power is not a single-valued function of distance to the shower axis. From \cite{Buitink:2014eqa}.}
\label{Fig_radio}
\end{center}
\end{figure}

Radio detection of air-showers was first demonstrated in 1965 \cite{1965Natur.205..327J}, but took a long time to become established as a reliable detection method. Perceived advantages that drove the development of the radio detection of air showers include (a) unlike ground particle arrays, the radio signal provides a calorimetric measurement of the full shower, (b) unlike Cherenkov or fluorescence telescopes, radio detection provides near-100\% duty cycle, and (c) radio antennas are inexpensive compared to particle detectors, allowing economic construction of large-area detectors. The breakthrough resulted from the development of the theory of emission and corresponding simulation tools \cite{Huege:2016veh}, combined with wide-band digital signal recording and signal processing; detailed reviews of the status of the technique and of detection systems are given in \cite{Huege:2016veh,Schroder:2016hrv}. Antenna systems for radio detection are generally wide-band, typically focusing on the frequency range 30-100~MHz -- where environmental noise is at a manageable level.
Current installations \cite{Schroder:2016hrv} -- usually combined with particle air shower arrays to trigger and independently determine shower properties -- range from few antennas to several hundred for LOFAR \cite{Schellart:2013bba}, with antenna densities 10 per km$^2$ to over 1000/km$^2$ for LOFAR, and areas covered as large as 17 km$^2$ for AERA, the radio extension of Auger~\cite{Abreu:2012pi}. Antenna types include simple dipole antennas, aperiodic loaded loop antennas, butterfly antennas, logarithmic periodic dipole antennas and directional antennas. A low-noise amplifier is usually integrated in or near the antenna, feeding an analogue-to-digital converter. Analogue and digital filtering serves to suppress man-made or natural noise. Antenna response is ideally calibrated in the field, accounting for the characteristics of the ground that may influence antenna response. The shower direction is determined from the arrival of the radio signal at the individual stations, in the same manner as done for particle air shower arrays. The shower core position is reconstructed from intensity distribution on the ground and the shape of the conical section of the wavefront; in addition, the frequency spectrum (i.e. pulse width) and polarization can be used. Best precision is reached by iteratively comparing data with simulations, properly accounting for the azimuthal dependence of signals arising from the interference of geomagnetic and Askaryan components (Fig. \ref{Fig_radio},\cite{Buitink:2014eqa}). Most current installations primarily serve to develop and validate the technique, rather than as `production facilities' with the prime purpose of providing air shower data. Also, the term `imaging' has limited validity; most analysis techniques used to process radio data are closer to the techniques of particle shower arrays, than to radio astronomy techniques where beam-forming in phased arrays can provide real sky images. For ground-based radio arrays, the interferometric method that includes the full phase information was, however, applied by the LOPES (LOFAR Prototype Station) experiment, reconstruction the radio sky map representing an extensive air shower \cite{Falcke:2005tc}. A complementary approach to ground radio arrays is followed by ANITA, with 32 horn antennas flying on a high-altitude balloon over Antarctica, measuring radio emission in the 200-1200 MHz band, primarily to search to signal create by ultra high energy neutrinos interacting in the ice, but also detecting radio emission by ultra high energy air showers, where the radio emission is reflected off the ice surface into the receivers \cite{Schoorlemmer:2015afa}.

Measurement of the distribution and polarization of radio signals from air showers above $10^{17}$ eV  \cite{Schroder:2016hrv} have confirmed and validated the results of simulation codes such as CoREAS \cite{Huege:2013vt}: (a) the polarization of radio signals and the relative strength of the Askaryan emission of 10\% to 20\% compared to the geomagnetic emission; (b) the radial distribution of the radio signal, falling off with a characteristic scale of 150~m to 200~m (see also Figs. \ref{fig_sho1} \ref{Fig_radio}); (c) the shape of the radio wave-front, approximately spherical close to the shower axis, and approaching a cone at distance of 50~m to 100~m; (d) the quadratic relation between energy of the primary and radio power, resulting from the coherent emission. The total energy radiated in the 30 to 80 MHz range is measured as $E_{30-80 \rm MHz} = (15.8 \pm 0.7 \pm 6.7) \mbox{MeV} (\sin\alpha\,E_{18} B_{\rm Earth}/0.24\,\mbox{G})^2$ where $E_{18}$ is the energy in $10^{18}$ eV and B the geomagnetic field \cite{2016PhRvD..93l2005A}, at an angle $\alpha$ to the shower axis; the Askaryan contribution is neglected in this parameterisation.
From fits to the wave-front, the direction of the primary was reconstructed with a precision of better than $1^\circ$; from the intensity distribution on the ground, the depth of shower maximum can be reconstructed with a precision of 20~g/cm$^2$. Predicted and measured radio intensity agree within the systematic uncertainty of $\approx 20 \%$. Towards large radio-based air shower detectors, the fact that emission is beamed resulting in a footprint similar to emission of Cherenkov light implies antenna spacing on the scale of 100\,m, at least if thresholds in the $10^{16} - 10^{17}$ eV domain are targeted. Rather than serving as stand-alone systems, upcoming radio installations are frequently considered in combination with particle arrays, in particular enhancing the determination of shower energy and of height of shower maximum, with applications for example in studies of cosmic ray composition.

\section{Underground Neutrino Detectors}

Whilst neutrino detection using air showers is promising at ultra-high
energies, in the TeV range there is no alternative to a shielded
instrumented volume. The TeV neutrino detectors are typical protected
from through-going muons by over a kilometer water or ice, with Cherenkov
light production in these media as the means of detection. Given the
large Cherenkov angle in these media ($\sim$40$^{\circ}$) detectors
need to be placed around the emitting particles, rather than `ahead'
as in the case of air-Cherenkov detectors. In any case, volume rather
than surface detectors are required as the interaction probability in
any plausible detector remains much lower than one up to ultra-high
energies. The required volume $V$ is set by the astrophysical neutrino flux ($F_\nu$, see Fig.~\ref{fig_cr}) and mean free path ($\lambda_\nu$, see Sec.~\ref{sec_det}): $V > \lambda_\nu / F_\nu t_{\mathrm{obs}}$, which for observation time $t{\mathrm{obs}}$ equal to one year is close to a cubic kilometre around 1~PeV for an ice or water target. The IceCube detector~\cite{2014ARNPS..64..101G} is the first detector constructed on this scale. For atmospheric neutrino detection at TeV energies the corresponding volume is of order $10^{6}$ m$^{3}$ or 1~Megatonne target mass, and a number of instruments on this scale now exist.

Below we discuss first general considerations for detection and reconstruction of different neutrino flavours, and then technical considerations on the design of experiments and individual detection elements.
%
%

\subsection{Detection Approach}

\begin{figure}[h]
\begin{center}
\includegraphics[width=0.48\textwidth, trim=30 55 220 30, clip]
{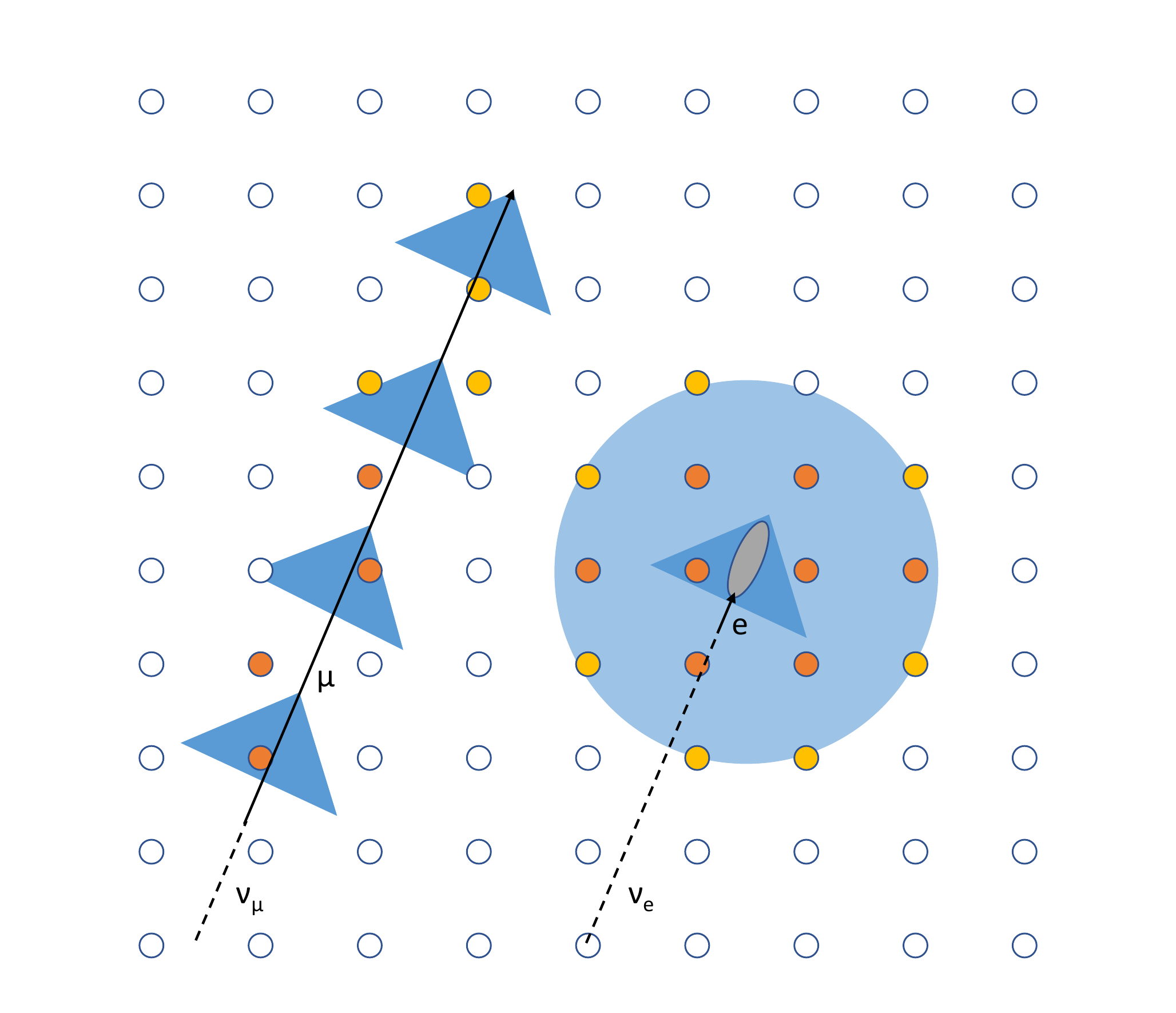}
\caption{Schematic representation of the detection of muon and electron neutrinos via the Cherenkov emission of thier interaction products. Left: a muon-neutrino interaction leads to the production of muon with a long track length in the ice. Cherenkov light is produced along the track and the orange coloured detectors recording earlier photon hits than those in yellow. Right: in the case of an electron-neutrino interaction the electron produced initiates an electomagnetic cascade which produces a large amount of Cherenkov light in a small volume, whilst the light is initially directed at the Cherenkov angle, scattering allows the light to be detected within a large volume (for a sufficiently energetic incoming particle), but with arrival time information giving only a weak indication of the incoming particle direction.
  }
\label{fig_trackcascade}
\end{center}
\end{figure}

For most source scenarios all three neutrino flavours arrive at the Earth in similar numbers and electrons, muons and taus generated in neutrino interactions all have distinguishable characteristics within the detector. The most straight-forward case are muons, which have km-scale path lengths in water or ice at TeV energies, increasing the effective volume of the detector and making precise direction reconstruction possible. For the same reason, even a deep underground detector has a residual background of down-going TeV muons from air-showers. Electrons by contrast initiate cascades which die out on the scale of tens of radiation lengths (metres).
Fig.~\ref{fig_trackcascade} illustrates the detection approach for muon and electron neutrinos in a high energy detector. The case of tau-neutrinos is particularly interesting experimentally because the generated tau lepton may be accompanied by a hadronic cascade and has a relatively long, $\approx50(E/\mathrm{1\,PeV})$ m, decay length and in 82\% of cases produces a cascade on decaying. This results in the possibility of a `double-bang' signature that unambiguous signals that the incoming neutrino was a $\nu_{\tau}$ and allows reasonable directional sensitivity.  

In the case of cascades the signal is proportional to the energy of the initiating EM particle and energy resolution reaches $\sim$20\% for well measured events. Below TeV energies muon energy losses are dominated by ionisation and the track length is proportional to the muons original energy: $l_\mu \sim 5 E_{\rm GeV}$ metres. At higher energies radiative process dominate and losses are proportional to muon energy, with a characteristic loss length of $\sim$ 5~km. In either regime muon energy estimation is therefore possible but is typically much worse than for cascades for non-contained events. 

In general searches are focused on up-going to horizontal events to discriminate against the residual background from down-going muons. However, at hundreds of TeV the Earth becomes opaque to neutrinos and the background of air-shower muons can be reduced to the point that down-going neutrino identification is possible. In general the effective area of a neutrino telescope is a strongly increasing function of energy, due to the combination of increasing neutrino cross-section and detection efficiency (for example due to the increasing muon range), see e.g.~\cite{2009NuPhS.190..101M}. As a consequence the detection threshold energy is typically poorly defined.

Note that radio detection within these deep detectors is also possible for cascade events of sufficient energy, but most current concepts use shallow deployment depths, relying on the $\sim$km propagation distances of 0.1-1 GHz frequency emission in ice. Such detectors could cover the neutrino energy range between the large Cherenkov detectors and balloon-born instrumentation such as ANITA (see section~\ref{sec_imaging}).

\subsection{Detector Design Considerations}

\begin{figure}[htbp]
\begin{center}
\includegraphics[width=9.6cm]{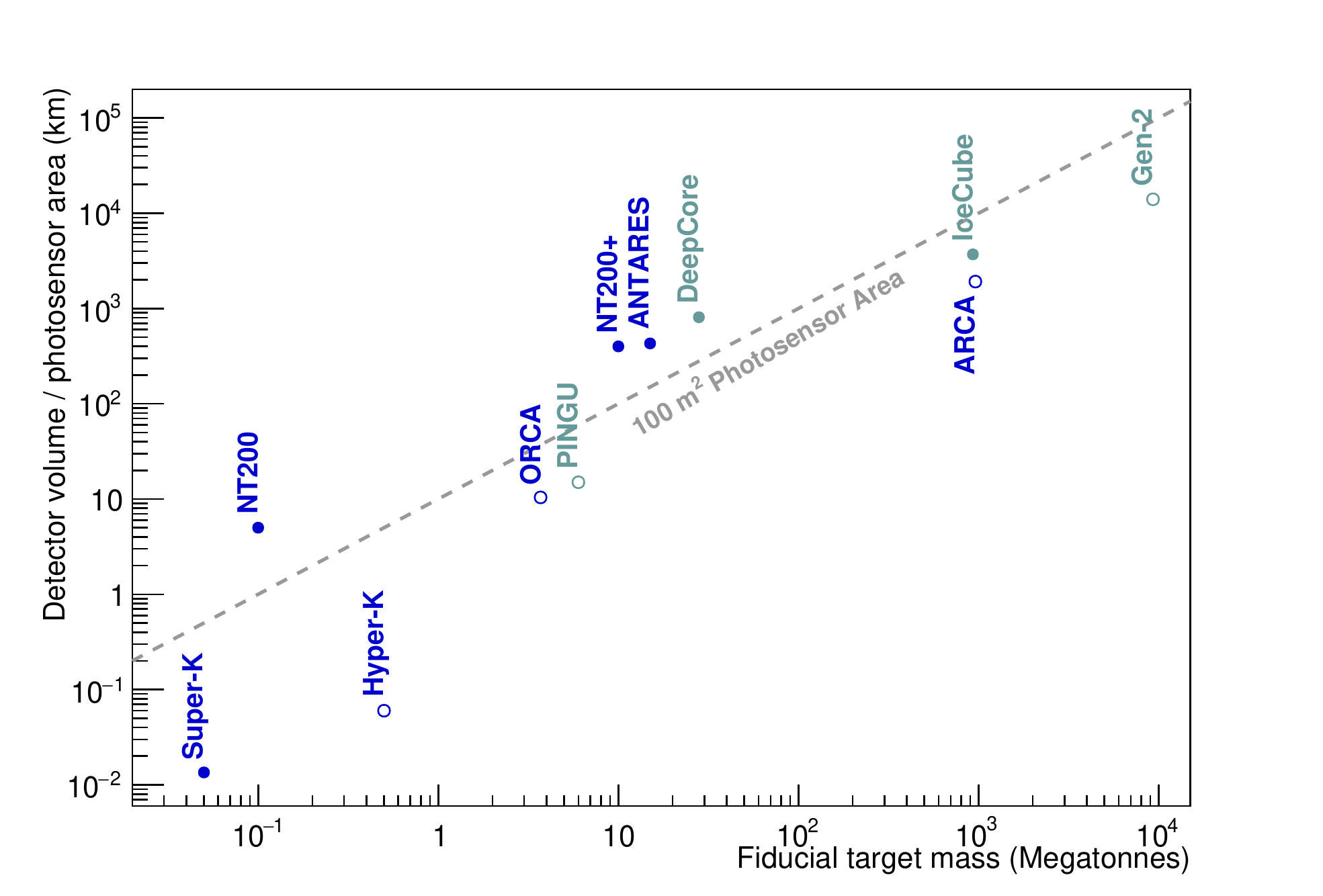}
\caption{Approximate photosensitive area divided by fiducial volume against (contained) target mass for a selection of neutrino detectors. Current or past detectors are marked with solid symbols, planned detectors with empty symbols. Water-based detectors have blue markers, those using ice as detection medium are shown in gray. The diagonal line indicates a constant photocathode area of 100~m$^{2}$. There is a very rough correspondence of the sensitive area/target volume plotted to the threshold energy of the detector, with MeV detectors at the bottom, GeV in the centre and TeV towards the top. NT200 and NT200+ are instruments located within Lake Baikal~\cite{2008NIMPA.588...99A}, ORCA and ARCA are part of the KM3Net programme~\cite{2016JPhG...43h4001A} and Gen-2 and PINGU are projects to extend the capabilities of IceCube at the South Pole~\cite{2016arXiv160702671T}\cite{2014arXiv1412.5106I}. 
}
\label{fig_nudets}
\end{center}
\end{figure}

Fig.~\ref{fig_nudets} illustrates the huge range of target mass of existing and planned neutrino detectors and the even larger range of photosensitive area required per unit of observed volume. At the extreme left edge of this plot are detectors in which a volume is observed from all sides by a close to fully instrumented surface.
The limiting scale of such a detector is set by the absorption and/or scattering lengths of photons in the detection medium.
In sea-water absorption lengths of $\sim$60~m have been measured, with scattering lengths of $\sim$300~m. In polar ice the situation is the opposite with scattering lengths around 30~m and absorption lengths around 100~m.

The Hyper-K project~\cite{2016NIMPA.824..630H}, with linear scale of $\sim$50~m is therefore approaching the limits for a detector of this type. The alternative is a sparsely filled volume, with light detectors deployed on a grid of `strings' due to obvious practicalities of the deployment process. The spacing of strings and of the individual light detectors on a string should normally not exceed the scattering or absorption scales by a significant factor. The Cherenkov yield ($\sim$ a few hundred photons per cm of track length) is such that of order 0.1~m$^{2}$ of detection area (at typical photon detection or quantum efficiency) is needed for detection of a single charged particle at a distance of some tens of metres.
This sensitive area must be housed in an enclosure designed to withstand extreme pressures, such an assembly is typically referred to as an optical modules (OM).
Modern detectors typically digitise the signal within the OM to avoid analogue transmission over km distances and associated loss of signal integrity. The simplest such device contains a single hemispheric PMT which must then be oriented up or down, with some loss of detection power in both cases (see above). A recent development are OMs housing a large number of smaller PMTs and providing close to uniform sensitivity in all directions~\cite{2015ICRC...34.1157B}.

Beyond scattering and absorption length differences, the alternative detection media for large detectors offer a range of different challenges in terms of deployment and operation. A sea-based detector must clearly be placed under a very large number of absorption lengths of water such that daytime operation is possible, and site selection is influenced by sea-bed data connection logistics and bio-luminescence activity. Typical single-photoelectron background rates are of the order of 50~kHz in sea-water, dominated by $^{40}$K decays but with the potentially time-variable component due to bio-luminescence. Only the antarctic ice cap so far appears suitable for large detectors, with large volumes of clear ice between dusty layers. In ice the deployment is based on hot-water drilling, refreezing after deployment creates pressure and changes ice properties close to the OM. Dark rates in ice are typically much lower than in water. In both media reference light sources are required to establish the wavelength-dependent scattering and absorption characteristics. For water-based detectors string motion is possible and some means of monitoring the detector geometry is needed. The near future plans for major instrumentation in the Mediterranean Sea are described in~\cite{2016JPhG...43h4001A} and for the South Pole in~\cite{2016arXiv160702671T} and~\cite{2014arXiv1412.5106I}. 

The precision of light-level measurements at an OM are not normally the main driver of performance: reconstruction usually makes use of a large number of low amplitude (single or a few photoelectron) measurements. More critical is the time resolution of an OM which should be on the order of the light crossing time of the photosensor ($\sim$1~ns), somewhat relaxed if scattering is relevant.

\section{Concluding Remarks}

Detectors for high energy messengers are playing an increasing important role in understanding the universe and in searching for new physics. Progress has been made by means of many novel detection techniques, by adopting techniques and technologies from particle physics and by the huge ambition and drive of the participants. The very brief overview given here hopefully provides an entry point to those unfamiliar with the field and something of interest for the experts.

{\bf Acknowledgements:} The authors thank K. Bernl\"ohr, R. Marx, H. Schoorlemmer, and F. Werner, for their help in preparing figures for this paper, and for reviewing the manuscript.


\bibliography{messengerbib}{}
\bibliographystyle{unsrt}

\end{document}